\documentclass[amsmath,amssymb,aps,pra,preprint,groupedaddress,longbibliography,superscriptaddress]{revtex4-1}
\usepackage{graphicx}% Include figure files
\usepackage{color}
\usepackage[colorlinks=true,urlcolor=blue,citecolor=blue,linkcolor=blue]{hyperref}
\newcommand{\bk}{\textbf{k}}
\newcommand{\bq}{\textbf{q}}
\newcommand{\bqt}{\tilde{\bf{q}}}

\begin{document}

% Use the \preprint command to place your local institutional report
% number in the upper righthand corner of the title page in preprint mode.
% Multiple \preprint commands are allowed.
% Use the 'preprintnumbers' class option to override journal defaults
% to display numbers if necessary
%\preprint{}

%Title of paper
\title{Superfluid weight and Berezinskii-Kosterlitz-Thouless temperature of spin-imbalanced and spin-orbit-coupled Fulde-Ferrell phases in lattice systems}

% repeat the \author .. \affiliation  etc. as needed
% \email, \thanks, \homepage, \altaffiliation all apply to the current
% author. Explanatory text should go in the []'s, actual e-mail
% address or url should go in the {}'s for \email and \homepage.
% Please use the appropriate macro foreach each type of information

% \affiliation command applies to all authors since the last
% \affiliation command. The \affiliation command should follow the
% other information
% \affiliation can be followed by \email, \homepage, \thanks as well.
\author{Aleksi Julku}
\author{Long Liang}
\author{P\"aivi T\"orm\"a}
%\email[]{Your e-mail address}
%\homepage[]{Your web page}
%\thanks{}
%\altaffiliation{}
\affiliation{Department of Applied Physics, Aalto University School of Science, FI-00076 Aalto, Finland}

%Collaboration name if desired (requires use of superscriptaddress
%option in \documentclass). \noaffiliation is required (may also be
%used with the \author command).
%\collaboration can be followed by \email, \homepage, \thanks as well.
%\collaboration{}
%\noaffiliation

\date{\today}

\begin{abstract}
We study the superfluid weight $D^s$ and Berezinskii-Kosterlitz-Thouless (BKT) transition temperatures $T_{BKT}$ in case of exotic Fulde-Ferrell (FF) superfluid states in lattice systems. We consider  spin-imbalanced systems with and without spin-orbit coupling (SOC) accompanied with in-plane Zeeman field. %In earlier studies FF phases have been predicted to exist in lattice geometries but the stability of these states against thermal phase flucutations has not been investigated before. 
By applying mean-field theory, we derive general equations for $D^s$ and $T_{BKT}$ in the presence of SOC and the Zeeman fields for 2D Fermi-Hubbard lattice models, and apply our results to a 2D square lattice. We show that conventional spin-imbalanced FF states without SOC can be observed at finite temperatures and that FF phases are further stabilized against thermal fluctuations by introducing SOC. We also propose how topologically non-trivial  SOC-induced FF phases could be identified experimentally by studying the total density profiles. Furthermore, the relative behavior of transverse and longitudinal superfluid weight components and the role of the geometric superfluid contribution are discussed.
\end{abstract}

% insert suggested PACS numbers in braces on next line
% insert suggested keywords - APS authors don't need to do this
%\keywords{}

%\maketitle must follow title, authors, abstract, \pacs, and \keywords
\maketitle

% body of paper here - Use proper section commands
% References should be done using the \cite, \ref, and \label commands
\section{Introduction}

Fulde-Ferrell-Larkin-Ovchinnikov (FFLO) superfluid states, identified by finite center-of-mass Cooper pairing momenta \cite{ff:1964,larkin:1964}, have gained widespread interest since their existence was predicted in the 1960s \cite{casalbuoni:2004}. Traditionally, FFLO states are considered in the context of spin-imbalanced degenerate Fermi gases where finite momenta of condensed Cooper pairs originate from the mismatch between the Fermi surfaces of two pairing Fermion species \cite{radzihovsky:2010,kinnunen:2018}. In such spin-polarized systems magnetism and superfluidity, usually thought to be incompatible with each other, co-exist and the superfluid order parameter is spatially varying, in contrast to the conventional Bardeen-Cooper-Schrieffer (BCS) pairing states characterized by the uniform order parameter and the absence of magnetism.

Realizing such spin-polarized FFLO states is challenging due to the requirement for large imbalance which in turn yields small superconducting order parameters and low critical temperatures. %However, recently it has been suggested \cite{zheng:2013,wu:2013,liu:2013,michaeli:2012,xu:2014,chen:2013,cao:2014,qu:2013,zhang:2013,huang:2017,dong:2013,hu:2013,zhou:2013,iskin:2013,iskin:2013b,wu:2013b} that FFLO states could exist in the presence of spin-orbit coupling (SOC) and Zeeman fields, for a review see \cite{kinnunen:2018}. 
In recent years, a very different physical mechanism for realizing FFLO phases, namely the introduction of spin-orbit coupling (SOC) and Zeeman fields, has been investigated in many theoretical studies \cite{zheng:2013,wu:2013,liu:2013,michaeli:2012,xu:2014,chen:2013,cao:2014,qu:2013,zhang:2013,huang:2017,dong:2013,hu:2013,zhou:2013,iskin:2013,iskin:2013b,wu:2013b,seo:2013,liu:2012,qu:2014,zheng:2014,guo:2018,guo:2017}, for a review see \cite{kinnunen:2018}. The advantage of these SOC-induced FFLO states is the absence of large spin polarizations as now finite Cooper pairing momenta originate from the deformation of the single-particle band dispersions and not from the mismatch of Fermi surfaces. As large polarizations are not needed, SOC-induced FFLO states might have higher critical temperatures than conventional imbalance-induced FFLO phases.

Despite many theoretical studies supporting the existence of FFLO phases, direct observation of such exotic superfluid states has been lacking \cite{casalbuoni:2004,beyer:2013}. For studying the FFLO state experimentally, ultracold Fermi gas systems are promising as they provide exact control of system parameters such as the spatial dimensionality, interaction strengths between the particles, and the system geometry \cite{esslinger:2010,bloch:2008,bloch:2012,torma:2014}. Ultracold gas experiments performed with quasi-one-dimensional population-imbalanced atomic gases have shown to be consistent with the existence of the FFLO state ~\cite{liao:2010} but unambiguous proof is still missing. 

In addition to conventional spin-imbalanced quantum gas experiments, recently also synthetic spin-orbit coupling and Zeeman fields have been realized in ultracold gas experiments \cite{lin:2011,wang:2012,cheuk:2012,zhang:2012,qu:2013_2} which makes it possible to investigate SOC-induced FFLO states as well. As SOC-induced FFLO states have been predicted to be stable in larger parameter regime than conventional spin-imbalanced FFLO phases \cite{xu:2014}, synthetic SOC could provide a way to realize FFLO experimentally in ultracold gas systems \cite{huang:2017}.

Low dimensionality has been predicted to favor FFLO-pairing \cite{parish:2007,koponen:2008}. However, in two and lower dimensional systems thermal phase fluctuations of the Cooper pair wave functions prevent the formation of true superfluid long-range order as stated by the Mermin-Wagner theorem \cite{mermin:1966}. Instead, only quasi-long range order is possible. In two dimensions, the phase transition from a normal Fermi gas to a superfluid state of quasi-long range order is determined by the Berezinskii-Kosterlitz-Thouless  (BKT) transition temperature $T_{BKT}$ \cite{kosterlitz:1973}. Below $T_{BKT}$ the system is a superfluid and above $T_{BKT}$ superfluidity is lost.

In recent years, SOC-induced FFLO phases in two-dimensional systems have gained considerable attention \cite{qu:2013,zhang:2013,xu:2014,wu:2013,zheng:2014,iskin:2013b,wu:2013b}. In these systems it has been argued that SOC accompanied with the in-plane Zeeman field would yield FFLO states. Furthermore, in ~\cite{qu:2013,zhang:2013} it was predicted that in the presence of the out-of-plane Zeeman field, i.e. spin-imbalance, SOC-induced FFLO states could be topologically non-trivial and support Majorana fermions. Such topological FFLO states are conceptually new and exotic superconductive phases of matter. However, these studies were performed by applying mean-field theories which do not consider the stability of FFLO states against thermal phase fluctuations in terms of the BKT transition. Superfluidity and BKT transition temperatures of BCS phases in spin-orbit-coupled Fermi gases have been theoretically investigated previously in \cite{lianyi:2012,gong:2012,devreese:2014,rosenberg:2017} but BKT transitions of  FFLO states have remained largely unstudied. As an exception, $T_{BKT}$ for FFLO states in case of a 2D \textit{continuum} system was explicitly computed in  \cite{yin:2014,cao:2014,cao:2015,xu:2015} where it was shown that SOC is required in order to have a non-zero $T_{BKT}$ for FFLO states. However, in case of spin-orbit coupled \textit{lattice} systems, $T_{BKT}$ of FFLO phases has not been studied before. Lattice systems are interesting since, due to Fermi surface nesting effects, the FFLO states are expected to be more stable and accessible than in continuum \cite{kinnunen:2018,koponen:2007,koponen:2008}. 

FFLO pairing states can be classified to two main categories: Fulde-Ferrell (FF) and Larkin-Ovchinnikov (LO) phases. In case of FF, the Cooper pair wave function $\Delta(\textbf{r})$ is a plane wave associated with a single pairing momentum so that it has a uniform amplitude but a spatially oscillating complex phase. The LO wave function, on the contrary, consists of two plane waves of opposite momenta and therefore has spatially varying amplitude. In spin-imbalanced systems without SOC, it has been shown, at the mean-field level, that in a square lattice the LO states should be slightly more energetically favorable than FF states \cite{baarsma:2016}, whereas in the presence of SOC both FF and LO states can exist as was shown in \cite{xu:2014}. Moreover, in \cite{guo:2018,guo:2017,iskin:2013b} the existence of topologically non-trivial FFLO phases in square and triangular lattices  was predicted. However, studies presented in \cite{xu:2014,guo:2018,guo:2017,iskin:2013b} did not consider the stability of  FFLO phases against thermal phase fluctuations.

In this work we investigate the stability of FF phases in lattice systems with and without SOC by calculating the BKT transition temperature $T_{BKT}$.  For a superconducting system the BKT temperature depends on the superfluid weight $D^s$ which is responsible for the dissipationless electric current and the Meissner effect - the fundamental properties of superconductors \cite{scalapino:1992,scalapino:1993}. In our study we develop a general theory for obtaining $D^s$ in any kind of lattice geometry in the presence of SOC and Zeeman fields, and apply the theory to a square lattice. We show that FF states  in a square lattice indeed have a finite $T_{BKT}$ with and without SOC, which is of fundamental importance as well as  a prerequisite for their experimental observation. Topological FF states created by the interplay of SOC and Zeeman fields are identified with the Chern numbers $C = \{\pm 1,-2\}$, and we explain how different topological FF phases can be distinguished by investigating the momentum density profiles which are experimentally accessible quantities. Additionally, we compare the superfluid weight components in orthogonal spatial directions. We also compute the so-called geometric superfluid weight component which is just recently found new superfluid contribution that depends on the geometric properties of the single-particle Bloch functions \cite{peotta:2015,liang:2017}.

%and compute the geometric superfluid weight contribution in various parameter regions.

In our study we discard the existence of LO phases as the LO ansatzes break the translational invariance which is required for deriving the superfluid weight in a simple form. Ignoring LO states, however, is not an issue because we are interested in the stability and BKT transition temperatures of exotic superfluid states: if there exists more stable LO states than FF states that we find, it implies the BKT transition temperatures of these LO states being higher than the temperatures we obtain for FF states. Therefore, our results can be considered as conservative estimates. Furthermore, in \cite{xu:2014,guo:2018} LO states were argued to exist when the superfluid pairing occurs within both helicity branches of a spin-orbit coupled square lattice. Thus, by studying the pairing amplitude profiles, we can deduce in which parts of our parameter space LO states would be more stable than the FF states we study.

The rest of the article is structured as follows. In the next section we provide expressions for the superfluid weight and thus for $T_{BKT}$ in the presence of SOC in case of an arbitrary lattice geometry. In section. \ref{section_three} we apply our equations for a spin-orbit coupled square lattice and show $T_{BKT}$ for various system parameters. We also discuss the topological properties of the system, and the different components of the superfluid weight. Lastly, in section \ref{section_four} we present concluding remarks and an outlook for future research. 

\section{Derivation of the superfluid weight in the presence of SOC for an arbitrary lattice geometry}\label{section_two}
% Put \label in argument of \section for cross-referencing
%\section{\label{}}
In this section we derive the expressions for the superfluid weight in the framework of BCS mean-field theory by applying linear response theory in a very similar way as was done in \cite{liang:2017}.  We consider the following two dimensional Fermi-Hubbard Hamiltonian
\begin{align}
H = &\sum_{i,j,\alpha,\beta,\sigma,\sigma'}t_{i\alpha\sigma,j\beta\sigma'}c^\dag_{i\alpha\sigma}c_{j\beta\sigma'} - \sum_{i\alpha\sigma}\mu_{\sigma}c^\dag_{i\alpha\sigma}c_{i\alpha\sigma} +U\sum_{i\alpha} c^\dag_{i\alpha\uparrow}c_{i\alpha\uparrow}c^\dag_{i\alpha\downarrow}c_{i\alpha\downarrow}
\end{align}
where $c^\dag_{i\alpha\sigma}$ creates a fermion in the $\alpha$-orbital of the $i$th unit cell with spin $\sigma \in \{\uparrow, \downarrow \}$. The first term describes the hopping processes which in addition to usual kinetic hopping terms ($\sigma = \sigma'$) can now also include spin-flipping terms ($\sigma \neq \sigma'$) required to take into account the spin-orbit coupling contribution. In the second term $\mu_\sigma$ is the spin-dependent chemical potential and the last term is the attractive on-site Hubbard interaction characterized by the coupling strength $U <0$. The above Hamiltonian describes any two-dimensional lattice geometry with arbitrary hopping and spin-flip terms, including the Rashba spin-orbit coupled two-component Fermi gases considered in this work.

We treat the interaction term by performing the standard mean-field approximation $Uc^\dag_{i\alpha\uparrow}c_{i\alpha\uparrow}c^\dag_{i\alpha\downarrow}c_{i\alpha\downarrow} \approx \Delta_{i\alpha}c_{i\alpha\downarrow}c_{i\alpha\uparrow} + \Delta_{i\alpha}^\dag c^\dag_{i\alpha\uparrow}c^\dag_{i\alpha\downarrow}$ where $\Delta_{i\alpha} = U\langle c_{i\alpha\downarrow}c_{i\alpha\uparrow} \rangle$ is the superfluid order parameter or in other words the wavefunction of the condensed Cooper pairs. To investigate the properties of the usual BCS and exotic inhomogeneous Fulde-Ferrell superfluid phases, we let the order parameter to have the form $\Delta_{i\alpha} = \Delta_{\alpha}\exp[i \tilde{\textbf{q}} \cdot \textbf{r}_i]$, where $\tilde{\textbf{q}}$ is the Cooper-pair momentum and $\textbf{r}_i$ is the spatial coordinate of the $i$th unit cell. The momentum of the Cooper pairs in a FF phase is finite, in contrast to a normal BCS phase where the Cooper pairs do not carry momentum.

By performing the Fourier transform to the momentum space $c_{i\alpha\sigma} = (1/\sqrt{N}) \sum_\bk e^{i\textbf{k}\cdot \textbf{r}_i}c_{\sigma\textbf{k}\alpha}$, where $N$ is the number of unit cells, one can rewrite the Hamiltonian in the form (discarding the constant terms)
\begin{align}\label{mfham}
H = \sum_\bk \Big ( & 
\begin{bmatrix}
c_{\uparrow \bk}^\dag & c_{\downarrow \bk}^\dag
\end{bmatrix}
\begin{bmatrix}
	\mathcal{H}_\uparrow(\bk) - \mu_\uparrow & \Lambda(\bk) \\
	\Lambda^\dag(\bk) & \mathcal{H}_\downarrow (\bk) - \mu_\downarrow 
\end{bmatrix}
\begin{bmatrix}
	c_{\uparrow \bk} \\
	c_{\downarrow \bk}
\end{bmatrix}\nonumber \\
&+ c_{\uparrow \bk}^\dag \Delta c_{\downarrow \bqt-\bk}^\dag + c_{\downarrow \bqt-\bk} \Delta^\dag c_{\uparrow \bk}
\Big ),
\end{align}
where $c_{\sigma\bk}^\dag = [c_{\sigma\textbf{k}1}, c_{\sigma\textbf{k}2},...,c_{\sigma\textbf{k}M}]$ and $\Delta = \textrm{diag}(\Delta_1,\Delta_2,...,\Delta_M)$, $M$ being the number of orbitals within a unit cell. Furthermore, $\mathcal{H}_\sigma(\bk)$ and $\Lambda(\bk)$  are the Fourier transforms of the kinetic hopping and the spin-flip terms, respectively.  

To write our Hamiltonian in a more compact form, let us introduce a four-component spinor $\Psi_\bk$ and rewrite the Hamiltonian as follows:
\begin{equation}\label{ham1}
H = \frac{1}{2}\sum_\bk \Psi^\dag_\bk \mathcal{H}_\bk \Psi_\bk,
\end{equation}
where
\begin{align}
\label{psispinor}
&\Psi_\bk = \begin{bmatrix}
c_{\uparrow \bk} \\
c_{\downarrow \bk} \\
c_{\downarrow \bqt - \bk}^\dag \\
-c_{\uparrow \bqt - \bk}^\dag
\end{bmatrix}
\equiv
\begin{bmatrix}
\psi_\bk \\
i\tau^y (\psi^\dag_{\bqt - \bk})^T
\end{bmatrix}
\equiv 
\begin{bmatrix}
\psi_\bk \\
\psi_{2,\bk}
\end{bmatrix}, \\
\label{hamp}
&\mathcal{H}_\bk = 
\begin{bmatrix}
\mathcal{H}_p(\bk) -\tilde{\mu} & \tilde{\Delta} \\
\tilde{\Delta}^\dag & - \mathcal{H}_h (\bk - \bqt) + \tau^y \tilde{\mu} \tau^y
\end{bmatrix}, \\
&\mathcal{H}_p(\bk) = \begin{bmatrix}
	\mathcal{H}_\uparrow(\bk) & \Lambda(\bk) \\
	\Lambda^\dag(\bk) & \mathcal{H}_\downarrow (\bk) 
\end{bmatrix},
\\
&\mathcal{H}_h(\bk) = -i \tau^y \mathcal{H}_p^*(-\bk)i \tau^y,
\\
&\tilde{\Delta} = \begin{bmatrix}
\Delta & 0 \\ 0 & \Delta
\end{bmatrix},
\\
&\tilde{\mu} = \begin{bmatrix}
\mu_\uparrow I_M & 0 \\ 0 & \mu_\downarrow I_M
\end{bmatrix}.
\end{align}
Here $\tau^y = \hat{\sigma}_y \otimes I_M$, where  $I_M$ is a $M\times M$ identity matrix and $\hat{\sigma} = [\hat{\sigma}_x,\hat{\sigma}_y,\hat{\sigma}_z]$ are the Pauli matrices. One should note that now the single-particle Hamiltonian is not anymore simply $\mathcal{H}_\sigma$ but $\mathcal{H}_p$ in which the two spin components are coupled via $\Lambda (\bk)$.

In two dimensions the total superfluid weight $D^s$ is a $2\times2$ tensor which reads 
\begin{align}
D^s = \begin{bmatrix}
	D^s_{xx} & D^s_{xy} \\
	D^s_{yx} & D^s_{yy}
\end{bmatrix},
\end{align}
where $x$ and $y$ are the spatial dimensions. To compute the superfluid weight tensor elements $D^s_{\mu\nu}$, we exploit the fact that at the mean-field level $D^s_{\mu\nu}$ is the long-wavelength, zero-frequency limit of the current-current response function $K_{\mu\nu}$ ~\cite{scalapino:1993}, that is
\begin{align}
\label{Ds_cc_response}
D^s_{\mu\nu} &= \lim_{\bq \rightarrow 0}  \lim_{\omega \rightarrow 0} K_{\mu\nu}(\bq,\omega) \nonumber \\
&= \lim_{\bq \rightarrow 0}  \lim_{\omega \rightarrow 0} \Big[ \langle T_{\mu\nu} \rangle  - i\int_0^\infty dt e^{i\omega t} \langle [ j^p_\mu(\bq,t),j^p_\nu(-\bq,0)] \rangle \Big],
\end{align}
where $j^p(\bq)$ and $T$ are the paramagnetic and diamagnetic current operators, respectively. The current operators can be derived by applying the Peierls substitution to the single-particle Hamiltonian $\mathcal{H}_p$ such that the hopping elements, both kinetic and spin-flipping terms,  are modified by a phase factor of $\exp[-i \textbf{A}\cdot (\textbf{r}_j - \textbf{r}_i)]$ where $\textbf{A}$ is the vector potential.  By assuming the phase factor to be spatially slowly varying, we can expand the Hamiltonian up to second order in $A$ to obtain $H = j^p_\mu A_\mu + T_{\mu\nu} A_\mu A_\nu/2$. In our case the $\mu$-component of the paramagnetic and diamagnetic current operators can be cast in the form
\begin{align}
\label{paramagnetic}
j_\mu^p(\bq) &= \sum_\bk \psi_{\bk + \bq}^\dag \partial_\mu \mathcal{H}_p(\bk + \bq/2) \psi_\bk \nonumber \\ 
&= \sum_\bk \Psi_{\bk + \bq}^\dag \partial_\mu \mathcal{H}(\bk + \bq/2)P_+ \Psi_\bk
\end{align}
and
\begin{align}\label{diamagnetic}
T_{\mu\nu}(\bq) &= \sum_\bk \psi_\bk^\dag \partial_\mu \partial_\nu \mathcal{H}_p(\bk) \psi_\bk \nonumber \\
&= \sum_\bk \Psi_\bk^\dag \partial_\mu \partial_\nu \mathcal{H}(\bk)P_+ \Psi_\bk, 
\end{align}
where $P_+ = (I_{4M}+ \hat{\sigma}^z \otimes I_{2M})/2$ and more generally $P_\pm = (I_{4M}\pm \hat{\sigma}^z \otimes I_{2M})/2$. 

We are interested in computing the current-current response function $K_{\mu\nu}(\bq,\omega)$ which at the limit of $\bq \rightarrow 0$, $\omega=0$ yields the superfluid weight $D^s_{\mu \nu}$. To this end, we first define a Green's function $G(\tau,\bk) = - \langle T \Psi_\bk(\tau) \Psi^\dag_\bk(0) \rangle$. In the Matsubara frequency space this reads $G(i\omega_n,\bk) = 1/(i\omega_n - \mathcal{H}(\bk))$ which follows from the quadratic form of the Hamiltonian \eqref{ham1}. Now, the current operators \eqref{paramagnetic}-\eqref{diamagnetic}, the Green's function and the Hamiltonian all have the same structure as those for conventional BCS theory developed in \cite{liang:2017}. Thus one can compute, by applying the Matsubara formalism and analytic continuation, the current-current response function in a similar fashion as done in \cite{liang:2017}. One starts from \eqref{Ds_cc_response}, inserts the expressions \eqref{paramagnetic}-\eqref{diamagnetic} for the current operators, deploys the Matsubara formalism, applies the diagrammatic expansion up to first order diagrams and obtains
\begin{align}
\label{ds_int}
K_{\mu\nu}(\bq,i\omega_n) =& \frac{1}{\beta} \sum_\bk \sum_{\Omega_m} \textrm{Tr} \Big[\partial_\mu \partial_\nu \mathcal{H}(\bk) P_+ G(i\Omega_m,\bk) \nonumber \\ 
	&+\partial_\mu \mathcal{H} (\bk + \bq/2) P_+ G(i\omega_n+i\Omega_m,\bk+\bq) \nonumber \\&\times \partial_\nu \mathcal{H}(\bk + \bq/2) \hat{\gamma}_z G(i\Omega_m,\bk) \Big]. 
\end{align}
where $\beta = 1/k_B T$, $\hat{\gamma}_z = \hat{\sigma}_z \otimes I_{2M}$, and $\omega_n$ ($\Omega_m$) are bosonic (fermionic) Matsubara frequencies. From \eqref{ds_int} one eventually obtains (see appendix \ref{app:derivation}):
\begin{align}\label{sfw}
D^s_{\mu\nu} =& K_{\mu\nu}(\bq \rightarrow 0,0) \nonumber \\=& 2\sum_{\bk,i,j} \frac{n(E_{j,\bk}) - n(E_{i,\bk})}{E_{i,\bk} - E_{j,\bk}} \Big( \langle \phi_i(\bk)| \partial_\mu \mathcal{H}(\bk) P_+ | \phi_j(\bk) \rangle  \nonumber \\ &\times\langle \phi_j(\bk) | P_- \partial_\nu \mathcal{H}(\bk) | \phi_i(\bk) \rangle \Big), 
\end{align}
where $n(E)$ is the Fermi-Dirac distribution and $|\phi_i(\bk) \rangle$ are the eigenvectors of $\mathcal{H}(\bk)$ with the eigenvalues $E_{i,\bk}$. For $i=j$, the prefactor should be understood as $-\partial_{E_i}n(E_i)$, which vanishes at zero temperature if the quasi-particle spectrum is gapped. For gapless excitations, $-\partial_{E_i}n(E_i)$ gives finite contribution even at zero temperature. We have benchmarked our superfluid weight relation \eqref{sfw} to earlier studies as discussed in appendix \ref{app:benchmark}.

The BKT transition temperature $T_{BKT}$ can be obtained from the superfluid weight tensor by using the generalized KT-Nelson criterion \cite{nelson:1977} for the anisotropic superfluid \cite{cao:2014,xu:2015}:
\begin{align}
T_{BKT} = \frac{\pi}{8}\sqrt{\det[D^s(T_{BKT})]}.
\end{align}
In the computations presented in this work $D^s$ is at low temperatures nearly a constant and therefore we can safely use the following approximation
\begin{align}
\label{tbkt}
T_{BKT} \approx \frac{\pi}{8}\sqrt{\det[D^s(T=0)]}.
\end{align}

In ~\cite{peotta:2015,liang:2017} it was shown that in case of conventional BCS states the superfluid weight can be divided to two parts: the so-called conventional and geometric contributions, $D^s_{\mu \nu} = D^s_{\textrm{conv},\mu \nu} + D^s_{\textrm{geom},\mu \nu}$. The conventional superfluid term $D^s_{\textrm{conv},\mu \nu}$ depends only on the single-particle energy dispersion relations, whereas the geometric part $D^s_{\textrm{geom},\mu \nu}$ comprises the geometric properties of the Bloch functions. In a similar fashion than in \cite{liang:2017}, also in our case the superfluid weight can be split to conventional and geometric parts so that $D^s_{\textrm{conv},\mu \nu}$ is a function of the single-particle dispersions of $\mathcal{H}_p$ and $\mathcal{H}_h$, and correspondingly  $D^s_{\textrm{geom},\mu \nu}$ depends on the Bloch functions of $\mathcal{H}_p$ and $\mathcal{H}_h$. The separation of $D^s$  to $D^s_{\textrm{geom}}$ and $D^s_{\textrm{conv}}$ terms is shown in appendix \ref{app:a}.

\section{Rashba-spin-orbit-coupled fermions in a square lattice}\label{section_three}
The above expression \eqref{sfw} of the superfluid weight holds for an arbitrary multiband lattice system. Here we focus on the simplest possible case, namely the square lattice geometry where the so-called Rashba spin-orbit coupling is applied to induce Fulde-Ferrell phases. By computing the superfluid weight and thus the BKT transition temperature, one can investigate the stability of SOC-induced FF phases versus the conventional FF phases induced by the spin-imbalance. We start by writing the Hamiltonian in the form
\begin{align}
H =& -t\sum_{\langle i,j \rangle,\sigma} c_{i\sigma}^\dag c_{j\sigma} - \mu \sum_{i\sigma} c_{i\sigma}^\dag c_{i\sigma} + U \sum_i c_{i\uparrow}^\dag c_{i\uparrow} c_{i\downarrow}^\dag c_{i\downarrow} \nonumber \\
&+ H_{z,in} + H_{z,out} + H_{SOC},
\end{align}
where the first term is the usual nearest-neighbour hopping term (we discard the orbital indices as in a square lattice there is only one lattice site per unit cell). The last three terms are the in-plane Zeeman field, out-of-plane Zeeman field and the Rashba coupling, respectively. They are
\begin{align}
&H_{z,in} = h_x \sum_i c_i^\dag \hat{\sigma}_x c_i \\
&H_{z,out} = h_z \sum_i c_i^\dag \hat{\sigma}_z c_i \\
&H_{SOC} = i\lambda \sum_{\langle i,j \rangle} c_i^\dag(\textbf{d}_{ij} \times \hat{\sigma})_z c_j.
\end{align}
Here $\textbf{d}_{ij}$ is the unit vector connecting the nearest-neighbour sites $i$ and $j$,  $\hat{\sigma} = [\hat{\sigma}_x, \hat{\sigma}_y, \hat{\sigma}_z]^{T}$ are the Pauli matrices and $c_i = [c_{i\uparrow},c_{i\downarrow}]^T$. The out-of-plane Zeeman fields can be included to the spin-dependent chemical potentials by writing $\mu_\uparrow = \mu + h_z$ and $\mu_\downarrow = \mu - h_z$. Furthermore, due to the in-plane Zeeman field and the Rashba spin-flipping terms, $\Lambda(\bk)$ in \eqref{mfham} has the form $\Lambda(\bk) = h_x  - 2\lambda (\sin k_y + i \sin k_x)$. We determine the order parameter amplitude $\Delta$ and the Cooper pair momentum $\tilde{\bq}$ self-consistently by minimizing the grand canonical thermodynamic potential $\Omega(\Delta,\tilde{\bq}) = -k_B T \log[\textrm{Tr}(e^{-\beta H})]$ which in the mean-field framework at $T=0$ reads as
\begin{align}
\label{therm_omega}
\Omega_{\textrm{M.F.}} = - \frac{\Delta^2}{U} + \frac{1}{2}\sum_{\bk,\nu,\eta}  E^\eta_{\bk,\nu}\Theta(-E^\eta_{\bk,\nu}),
\end{align}
where $\Theta(x)$ is the Heaviside step function and $E^\eta_{\bk,\nu}$ are the eigenvalues of $\mathcal{H}_\bk$. Here $\eta = \{+,- \}$ labels the quasi-particle and quasi-hole branches, respectively and $\nu = \{1,2\}$ the helicity branches split by the spin-orbit coupling. The quasi-particle branches are taken to be the two highest eigenvalues of $\mathcal{H}_\bk$. In \eqref{therm_omega} we have discarded the constant term $\sum_\bk \textrm{Tr}[\mathcal{H}_h(\bk - \tilde{\bq}) - \tau^y\tilde{\mu} \tau^y]$ which is not needed when one minimizes $\Omega_{\textrm{M.F.}}$. Consistent with previous lattice studies \cite{xu:2014,guo:2018,guo:2017}, the Cooper pair momentum is in the $y$-direction, i.e. $\tilde{\bq} = \tilde{q}_y \hat{\textbf{e}}_y$ as the in-plane Zeeman field in the $x$-direction deforms the single-particle dispersions in the $y$-direction. We have numerically checked that the solutions with the Cooper pair momentum in the $y$-direction minimize the thermodynamic potential, as discussed in appendix \ref{app:qbm}. When the correct values for $\Delta$ and $\tilde{q}_y$ are found, the superfluid weight can be computed with  \eqref{sfw}. 

We investigate the topological properties by computing the Chern number $C$ for our interacting system by integrating the Berry curvature $\Gamma^\eta_\nu(\bk)$ associated with the quasi-hole branches $\eta = -$ over the first Brillouin zone  as follows:
\begin{align}
C &= \frac{1}{2\pi}\sum_{\nu=1}^2 \int_{-\pi}^\pi  \int_{-\pi}^\pi dk_x dk_y \Gamma^{-}_\nu(\bk).
\end{align}
The explicit form for the Berry curvature can be expressed with the eigenvalues $E^\eta_{\bk,\nu}$ of $\mathcal{H}_\bk$ and the corresponding eigenvectors $| n(\bk) \rangle$, where $n = (\eta,\nu)$,  in the form
\begin{align}
\Gamma^\eta_\nu(\bk) = i \sum_{n \neq n'}\frac{\langle n | \partial_{k_x} \mathcal{H}_\bk | n' \rangle \langle n' | \partial_{k_y} \mathcal{H}_\bk | n \rangle - (k_x \leftrightarrow k_y)}{\big( E^\eta_{\bk,\nu} - E^{\eta'}_{\bk,\nu'}  \big)^2}.
\end{align}

\section{Results}
\subsection{Phase diagrams and the BKT temperature}

By deploying our mean-field formalism we determine the phase diagrams and $T_{BKT}$ as functions of the Zeeman fields and the average chemical potential $\mu = (\mu_\uparrow + \mu_\downarrow)/2$. We fix the temperature to $T=0$ as, according to \eqref{tbkt}, the zero-temperature superfluid weight gives a good estimate for $T_{BKT}$. In all the computations we choose $t = 1$ and $U= -4$. Furthermore, we let $\tilde{q}_y$ to have only discrete values in the first  Brillouin zone such that $\tilde{q}_y \in \{\frac{\pi n}{L},n=1,2,...L \} $, where $L$ is the length of the lattice in one direction, i.e. the total number of lattice sites is $N = L \times L$. In all of our computations we choose $L=104$ and deploy periodic boundary conditions.

\begin{figure}
\includegraphics[width=0.75\columnwidth]{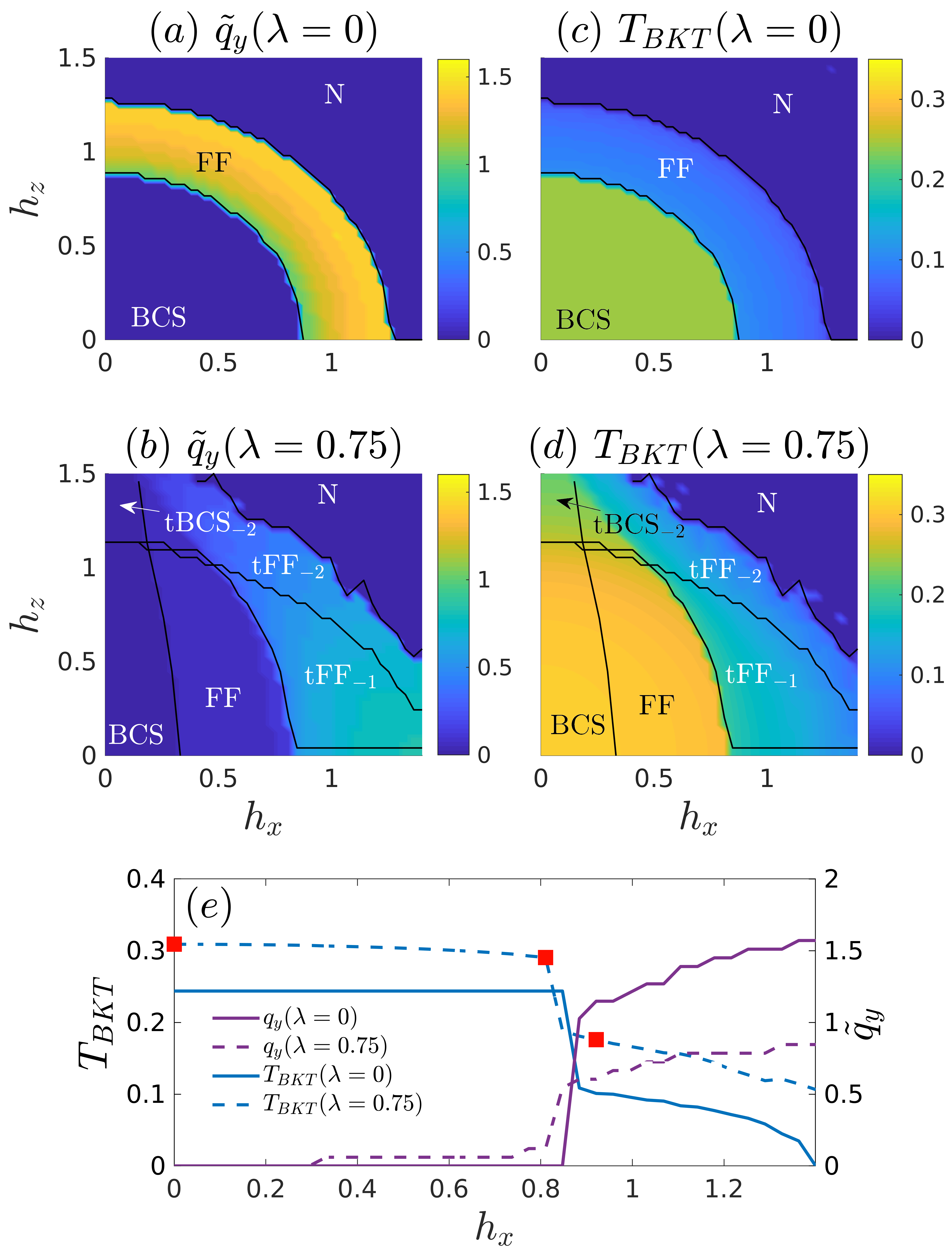}
\caption{\label{fig:1}(a)-(d) Cooper pair momentum $\tilde{q}_y$ and the corresponding BKT temperature $T_{BKT}$ as a function of the Zeeman fields $h_x$ and $h_z$ for the spin-orbit couplings $\lambda =0$ [(a) and (c)] and for $\lambda =0.75$ [(b) and (d)] at $\mu = 0.95$. In (a)-(b) the colors depict the magnitude of $\tilde{q}_y$ and in (c)-(d) the BKT temperature. For $\lambda =0$ all the phases are topologically trivial whereas for finite SOC there exists topologically non-trivial BCS and FF phases. Labels tFF$_{-1}$, tFF$_{-2}$ and tBCS$_{-2}$ correspond to topologically non-trivial FF and BCS phases of Chern numbers $-1$ and $-2$. In case of $\lambda = 0.75$ there exists two different FF regions, one with small Cooper pair momentum but large $T_{BKT}$ and one with larger $\tilde{q}_y$ but small $T_{BKT}$. (e) $T_{BKT}$ and $\tilde{q}_y$ as a function of $h_x$ at $h_z = 0$ for $\lambda = 0$ (purple lines) and $\lambda = 0.75$ (blue lines). Three red squares correspond to cases considered in figure \ref{fig:2}.}
\end{figure}

In figures~\ref{fig:1}(a)-(b) the superfluid phase diagrams in terms of the magnitude of $\tilde{q}_y$ are presented as a function of $h_x$ and $h_z$ at $\mu = 0.95$ for $\lambda = 0$ and $\lambda = 0.75$, respectively, and the corresponding BKT transition temperatures $T_{BKT}$ are shown in figures~\ref{fig:1}(c)-(d). From figure~\ref{fig:1}(a) we see that in the absence of SOC the phase diagram is symmetric with respect to the Zeeman field orientation. This is due to the SO(2) symmetry as under the rotation $\mathcal{U} [c_{i\uparrow},c_{i\downarrow}]^T \mathcal{U}^{-1} = \frac{1}{\sqrt{2}}[c_{i\uparrow} + c_{i\downarrow},c_{i\uparrow} - c_{i\downarrow}]^T \equiv [d_{i\uparrow}, d_{i\downarrow}]^T$ the Hamiltonian remains invariant except $h_x \rightarrow h_z$ and  $h_z \rightarrow h_x$. For small Zeeman fields, the BCS phase is the ground state and becomes only unstable against the FF phase for larger Zeeman field strengths. One can see from figure~\ref{fig:1}(c) that the BKT temperature for the BCS phase is $T_{BKT} \approx 0.25t$ and roughly $T_{BKT} \approx 0.1t$ for the FF phase. This implies that conventional imbalance-induced FF phases without SOC could be observed in lattice  systems, in contrast to continuum systems where it is shown that $T_{BKT} = 0$ \cite{yin:2014}. This is the first time that the stability against the thermal phase fluctuations of spin-imbalanced FF states in a lattice system is confirmed.

Unlike in the case of without SOC, the phase diagram shown in figure~\ref{fig:1}(b) for $\lambda = 0.75$ depends on the direction of the total Zeeman field, as SOC together with the in-plane Zeeman field breaks the $SO(2)$ symmetry. The interplay of the SOC and the Zeeman fields stabilize inhomogeneous superfluidity in larger parameter regions than in case of conventional spin-imbalanced FF states. Furthermore, by introducing SOC one is able realize topologically distinct BCS and FF phases. As with $\lambda =0$, at small Zeeman fields there exist topologically trivial BCS states. When $h_x$ is increased, the system enters non-topological FF phase and eventually for large enough $h_x$ topological FF states of $C = -1$  (tFF$_{-1}$) and $ C= -2$ (tFF$_{-2}$). By applying large $h_z$ one is able to reach topological BCS and FF phases, tBCS$_{-2}$ and tFF$_{-2}$, characterized by $C = -2$. For large enough Zeeman fields the superfluidity is lost and the system enters normal (N) state.

From figure~\ref{fig:1}(b) we see that in addition to topological classification, FF phases can be further distinguished by the magnitude of the Cooper pair momentum $\tilde{q}_y$: for intermediate Zeeman field strengths the FF state is characterized by rather small $\tilde{q}_y$, in contrast to region of large Zeeman fields where the pairing momenta are comparable to those of FF states of $\lambda = 0$. The same behavior can be seen by observing $T_{BKT}$ presented in figure~\ref{fig:1}(d). We see that for small-$\tilde{q}_y$ region $T_{BKT}$ is around $0.3t$ and becomes only smaller for large-$\tilde{q}_y$ region where $T_{BKT}$ at largest is roughly $T_{BKT} \approx 0.17t$. Therefore, by deploying SOC, one is able to stabilize FF phases considerably against thermal phase fluctuations and increase $T_{BKT}$. This is similar to continuum studies \cite{cao:2014,cao:2015,xu:2015} where it was proposed that FF states could be observed with the aid of SOC. The difference of $\lambda = 0$ and $\lambda = 0.75$ is further demonstrated in figure~\ref{fig:1}(e), where $T_{BKT}$ and $\tilde{q}_y$ for both the cases are plotted as a function of $h_x$ at $h_z = 0$. We see that the phase diagram becomes richer and $T_{BKT}$ is increased when SOC is deployed.

\begin{figure}
\includegraphics[width=1.0\columnwidth]{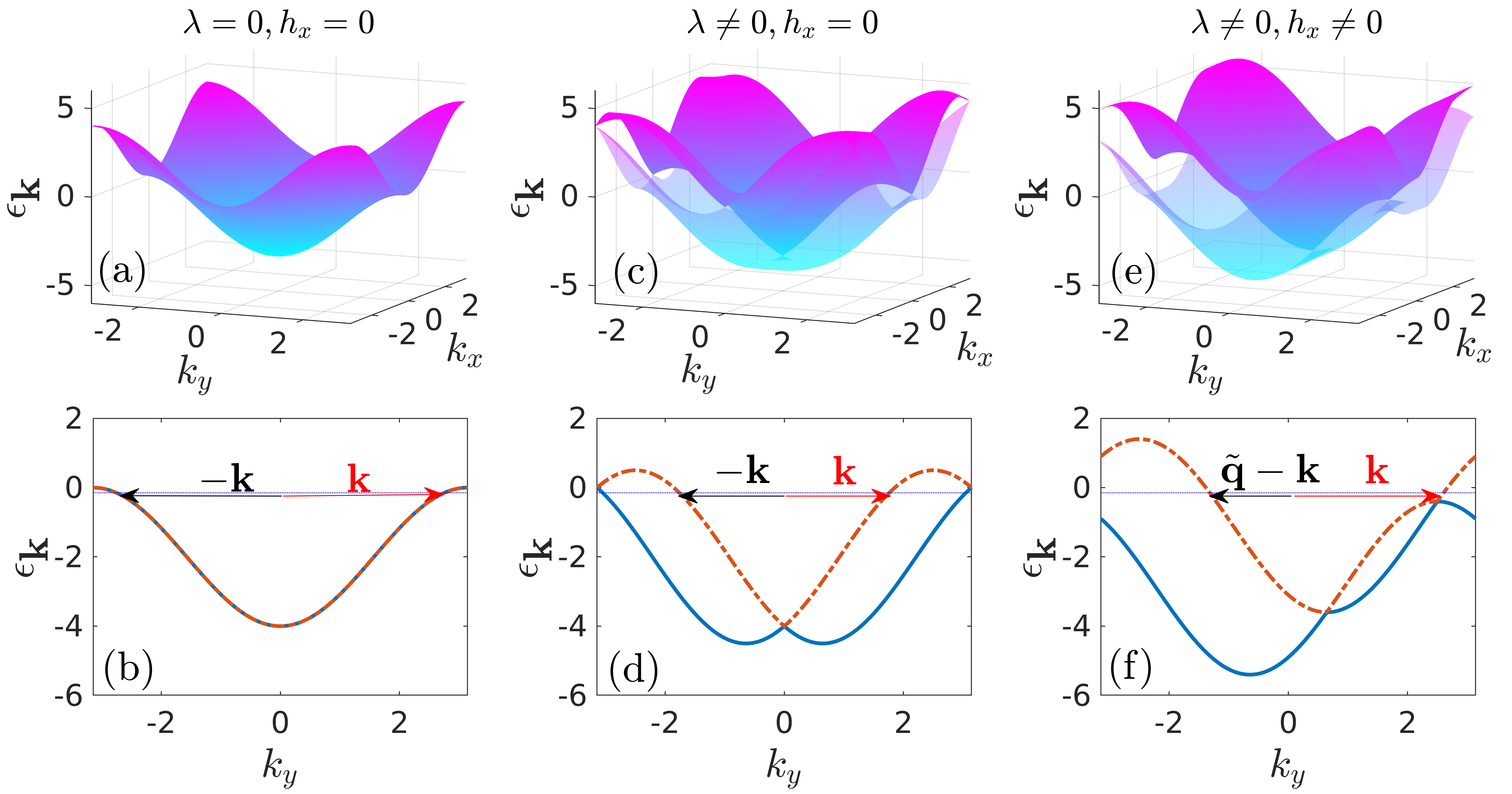}
\caption{\label{fig:extra}Schematics of single-particle dispersions in case of $\lambda =0$, $h_x = 0$ [(a)-(b)], $\lambda \neq 0$, $h_x = 0$ [(c)-(d)] and $\lambda \neq 0$, $h_x \neq 0$ [(e)-(f)]. The upper panels show the dispersions across the first Brillouin zone and the lower ones at $k_x = 0$. Finite SOC splits the degenerate spin-up and spin-down dispersions to two branches and finite $h_x$ deforms the dispersions non-symmetrically with respect to $k_y=0$. In the lower panels the solid blue and dash-dotted red lines depict the dispersions, the black and red arrows depict the intraband pairing momenta and the blue dotted lines the Fermi surfaces. Here only the pairing within one band is depicted but in general, depending on the Fermi level and the Zeeman fields, pairing within both bands can occur. In the presence of the interband pairing, the Cooper pair momentum can in general deviate from the $y$-direction.}
\end{figure}
\begin{figure}
\includegraphics[width=1.0\columnwidth]{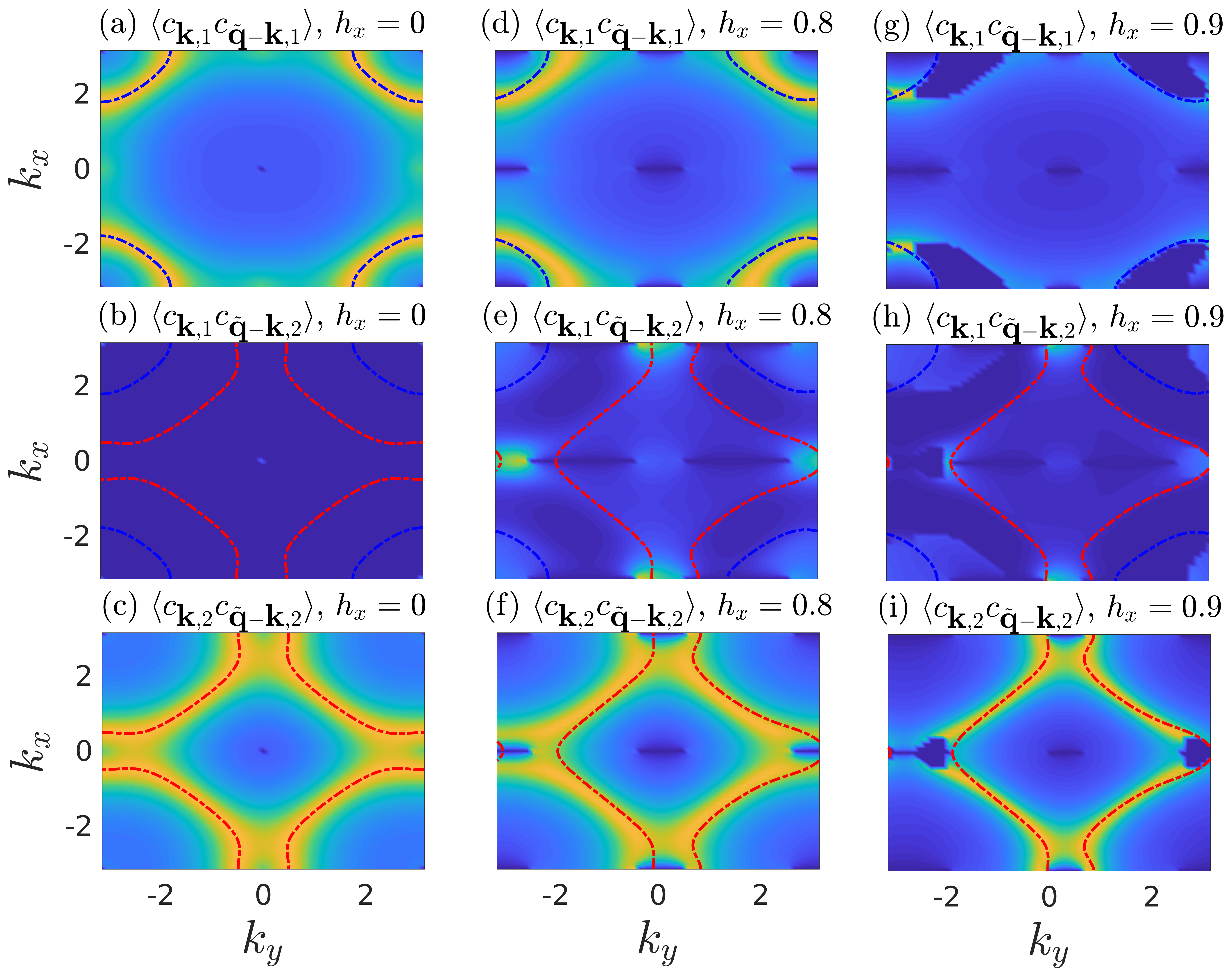}
\caption{\label{fig:2} Inter- and intraband pairing functions $|\langle c_{\bk,n} c_{\tilde{\bq} - \bk, n'} \rangle|$ for $h_x = 0$ [(a)-(c)], $h_x = 0.8$ [(d)-(f)] and $h_x = 0.9$ [(g)-(i)] in case of $\lambda = 0.75$, $\mu = 0.95$ and $h_z = 0$. These three cases correspond to the three red squares in figure \ref{fig:1}(e). The non-interacting Fermi surfaces are depicted as red (blue) contours for the upper (lower) dispersion band.}
\end{figure}

To understand why in the presence of SOC there exist distinct FF regions of considerably different BKT temperatures, we investigate the inter- and intraband pairing functions $\langle c_{\bk,n} c_{\tilde{\bq} - \bk, n'} \rangle$, where $ c_{\bk,n}$ is the annihilation operator for the $n$th Bloch function of the single-particle Hamiltonian $\mathcal{H}_p(\bk)$. In case of a square lattice, $\mathcal{H}_p(\bk)$ is a $2\times2$ matrix so we have two energy bands, called also helicity branches. As an example, in figure~\ref{fig:extra} the single-particle energy dispersion bands have been plotted at $h_z = 0$ for $\lambda =0$, $h_x = 0$ [figures~\ref{fig:extra}(a)-(b)], $\lambda \neq 0$, $h_x = 0$ [figures~\ref{fig:extra}(c)-(d)] and $\lambda \neq 0$, $h_x \neq 0$ [figures~\ref{fig:extra}(e)-(f)]. Without SOC, the single particle dispersions for spin up and down components are degenerate [figures~\ref{fig:extra}(a)-(b)]. By turning on the spin-orbit coupling, this degeneracy is lifted [figures~\ref{fig:extra}(c)-(d)] and when also $h_x$ is applied, the dispersion becomes deformed in a non-symmetric way with respect to $k_y = 0$ [figures~\ref{fig:extra}(e)-(f)]. This deformation of the dispersions results in the intraband pairing of finite momentum in the $y$-direction when $h_x$ is large enough as there exists a momentum mismatch of $\tilde{q}_y\hat{\textbf{e}}_y$ between the pairing fermions. If in addition the interband pairing occurs, the momentum mismatch can exist also in the $x$-direction and consequently the Cooper pair momentum is not necessarily in the $y$-direction. However, in the computations presented in this work $\tilde{\bq}$ has been numerically checked to be always in the $y$-direction.

With figures~\ref{fig:extra}(e)-(f) one can also understand the fundamental differences between conventional spin-imbalanced-induced and SOC-induced FF states in terms of spontaneously broken symmetries. Both cases break the time-reversal symmetry (TRS) spontaneously and in case of spin-imbalanced FF also the rotational symmetry within the lattice plane is spontaneously broken. In other words, for imbalance-induced FF states, it is energetically equally favorable for the Cooper pair momentum to be in the $x$- or $y$-direction. 
However, SOC and the in-plane Zeeman field break the rotational symmetry explicitly, and therefore the Cooper pair wavevector is forced to be in the perpendicular direction with respect to the in-plane Zeeman field as the dispersions are deformed in that direction [figures~\ref{fig:extra}(e)-(f)].

Even if the in-plane Zeeman field causes the single-particle dispersion to be non-centrosymmetric, it is still not a sufficient condition to reach the FF state as can be seen in figure~\ref{fig:1}(b) where the ground state is BCS for small enough values of $h_x$. Homogeneous BCS states can be still more favorable than FF states if for example the chemical potential is such that the shapes and the density of states of the Fermi surfaces prefer the Cooper pairing with zero momentum. However, when the in-plane Zeeman field becomes strong enough, the deformation of the dispersion results in the FF pairing.   

In figures \ref{fig:2}(a)-(i) we present $|\langle c_{\bk,1} c_{\tilde{\bq} -\bk,1} \rangle|$, $|\langle c_{\bk,1} c_{\tilde{\bq} -\bk,2} \rangle|$ and $|\langle c_{\bk,2} c_{\tilde{\bq} -\bk,2} \rangle|$ for $h_x = 0$ [(a)-(c)], $h_x = 0.8$ [(d)-(f)] and $h_x = 0.9$ [(g)-(i)] in case of $\lambda = 0.75$, $\mu = 0.95$ and $h_z = 0$. These three cases correspond to three red squares of figure \ref{fig:1}(e). For clarity, also the non-interacting Fermi surfaces are depicted as red (blue) contours for the upper (lower) branch. The case $h_x = 0$ shown in figures \ref{fig:2}(a)-(c) corresponds to conventional BCS phase for which intraband pairing takes place within both bands and interband pairing is vanishingly small. When $h_x$ is finite, the system enters first to the small-$\tilde{q}_y$ region [figures \ref{fig:2} (d)-(f)] where both intraband pairing contributions are still prominent and the interband pairing is finite but small. Due to the contribution of both bands, $T_{BKT}$ is more or less the same as for $h_x = 0$, see figure \ref{fig:1}(e). The only qualitative difference is the asymmetric pairing profiles of $h_x = 0.8$ which causes the finite momentum pairing to be more stable than the zero-momentum BCS pairing.

The situation is drastically different when the system enters to the large-$\tilde{q}_y$ region at $h_x = 0.9$ [figures \ref{fig:2} (g)-(i)]. In contrast to cases with smaller $h_x$, the prominent intraband pairing contribution comes now from the upper band alone. As the pairing occurs only in one of the bands instead of both bands, $T_{BKT}$ is significantly lower for the large-$\tilde{q}_y$ region than for the small-$\tilde{q}_y$ phase,  as seen in figure \ref{fig:1}(e).

It should be reminded that we consider FF states only and ignore LO states. In recent real-space mean field studies \cite{xu:2014,guo:2018}, it was pointed out that LO states are associated with  finite pairing amplitudes occurring within both bands and correspondingly FF phases are a consequence of the pairing occurring within a single helicity band only. This is easy to understand as the in-plane Zeeman field shifts the other helicity band to $+k_y$ and the other to $-k_y$ direction. Therefore, when the pairing occurs within both bands, some pairing occurs with Cooper pair momentum $+\tilde{q}_y$ and some with $-\tilde{q}_y$ which results in an LO phase. Thus, the small-$\tilde{q}_y$ region we find is likely the one where LO states are more stable than FF states and hence $T_{BKT}$  is considerably higher for LO states than for FF states. Unfortunately, accessing LO states directly is not possible with our momentum-space study as LO phases break the translational invariance which is utilized in the derivation of the superfluid weight as shown in section \ref{section_two}. For computing the superfluid weight also in case of LO ansatzes, one should derive the expressions for the superfluid weight by using real-space quantities only.

\begin{figure}
\includegraphics[width=1.0\columnwidth]{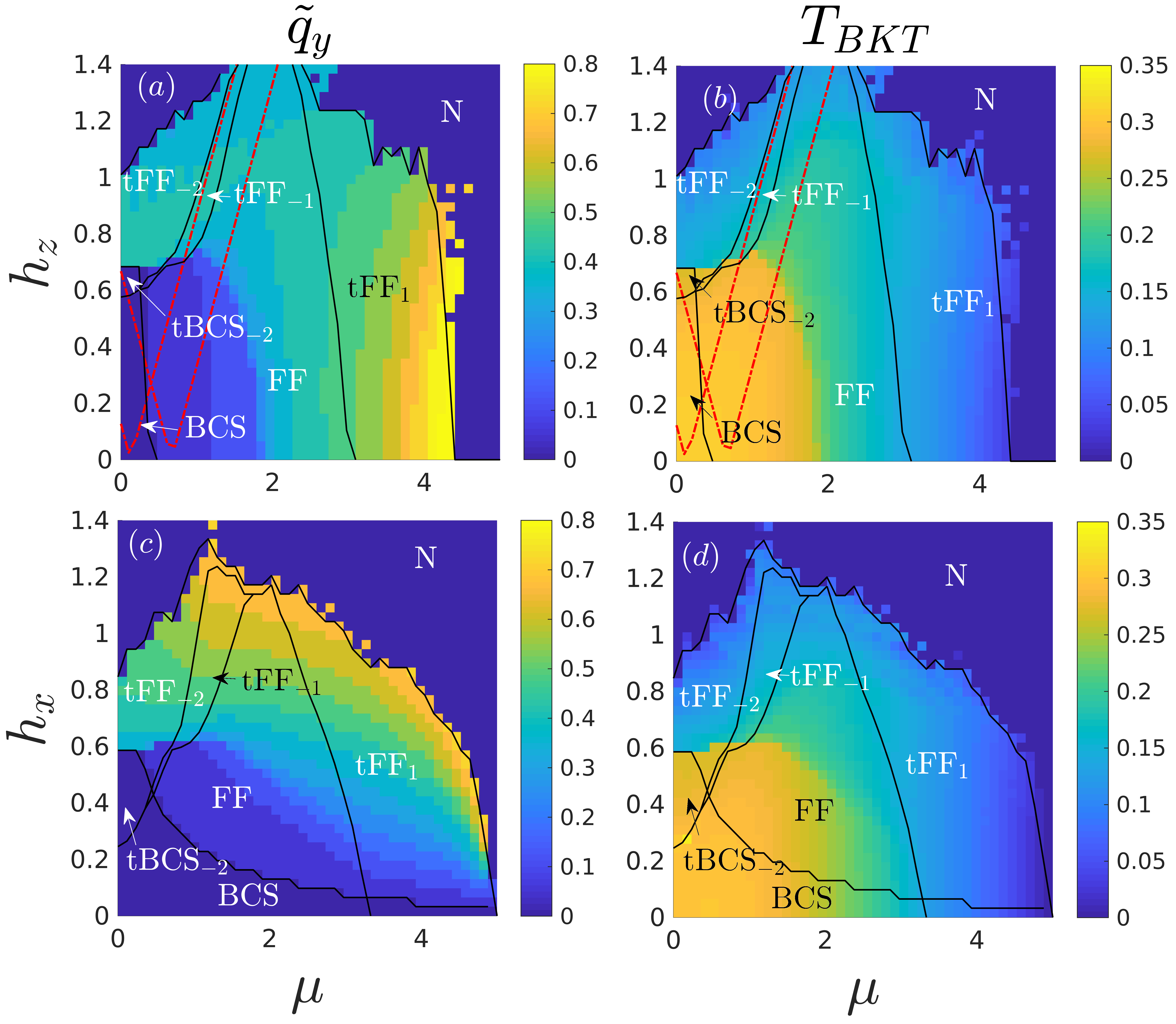}
\caption{\label{fig:3} Cooper pair momentum $\tilde{q}_y$ and the BKT temperature $T_{BKT}$ as a function of $\mu$ and $h_z$ [(a)-(b)] and as a function of $\mu$ and $h_x$ [(c)-(d)] for $\lambda = 0.75$. In (a)-(b) $h_x = 0.658$ and in (c)-(d) $h_z = 0.8$. Labels tFF$_{\pm 1}$, tFF$_{-2}$ and tBCS$_{-2}$ correspond to topologically non-trivial FF and BCS phases of Chern numbers $\pm 1$ and $-2$. Most stable FF phases are once again the ones identified by small Cooper pair momenta. As in figure \ref{fig:1}, also here we see various topological BCS and FF phases distinguished by different Chern numbers. The red dash-dotted line in (a)-(b) depict two of the Van Hove singularities of the square lattice system with spin-orbit coupled fermions.}
\end{figure}

For completeness, in figure~\ref{fig:3} we provide the phase diagrams for $\tilde{q}_y$ and $T_{BKT}$ as functions of $\mu$ and $h_z$ [figures \ref{fig:3}(a)-(b)] and of $\mu$ and $h_x$ [figures \ref{fig:3}(c)-(d)] at $\lambda = 0.75$. In case of the $(\mu,h_z)$-phase diagram the in-plane Zeeman field is fixed to $h_x = 0.658$ and in case of the $(\mu,h_x)$-diagram the out-of-plane Zeeman field is $h_z = 0.8$. As in figure~\ref{fig:1} with $(h_x,h_z)$-diagram, also here we find various topologically non-trivial FF and BCS phases identified with the Chern numbers $C = -1$ and $C=-2$ near the half-filling. However, for higher chemical potential values we also find topological FF and BCS phases characterized by $C = 1$. Furthermore, we can once again identify FF phases with high $T_{BKT}$ but considerably small Cooper pair momenta existing near the half filling with moderately low Zeeman field values. From figures~\ref{fig:3}(b) and (d) we see that for a non-topological FF phase $T_{BKT}$ is $0.1$-$0.3t$ at relatively large parameter regime. For topological FF states $T_{BKT}$ is somewhat lower, the maximum transition temperature being $T_{BKT} \sim 0.15t$.

In previous FFLO studies \cite{kinnunen:2018,koponen:2008,baarsma:2016} it has been shown that Van Hove singularities associated with the divergent behavior of the density of states near the Fermi surface can enlarge the parameter regime of FFLO states. In our spin-orbit-coupled square lattice system there are six different Van Hove singularities for fixed $\mu$. In figures~\ref{fig:3}(a)-(b) two of these singularities are depicted with red dash-dotted lines, the other four occurring near the depicted two. One can see that in the vicinity of the Van Hove singularities the FF phases can exist at higher values of $h_z$ than away from the singularities. However, in $(\mu,h_x)$-diagrams depicted in figures~\ref{fig:3}(c)-(d) the Van Hove singularities are not playing a role and therefore they are not shown.

%fig4_a-l_1203.jpg
\begin{figure*}
\includegraphics[width=0.92\textwidth]{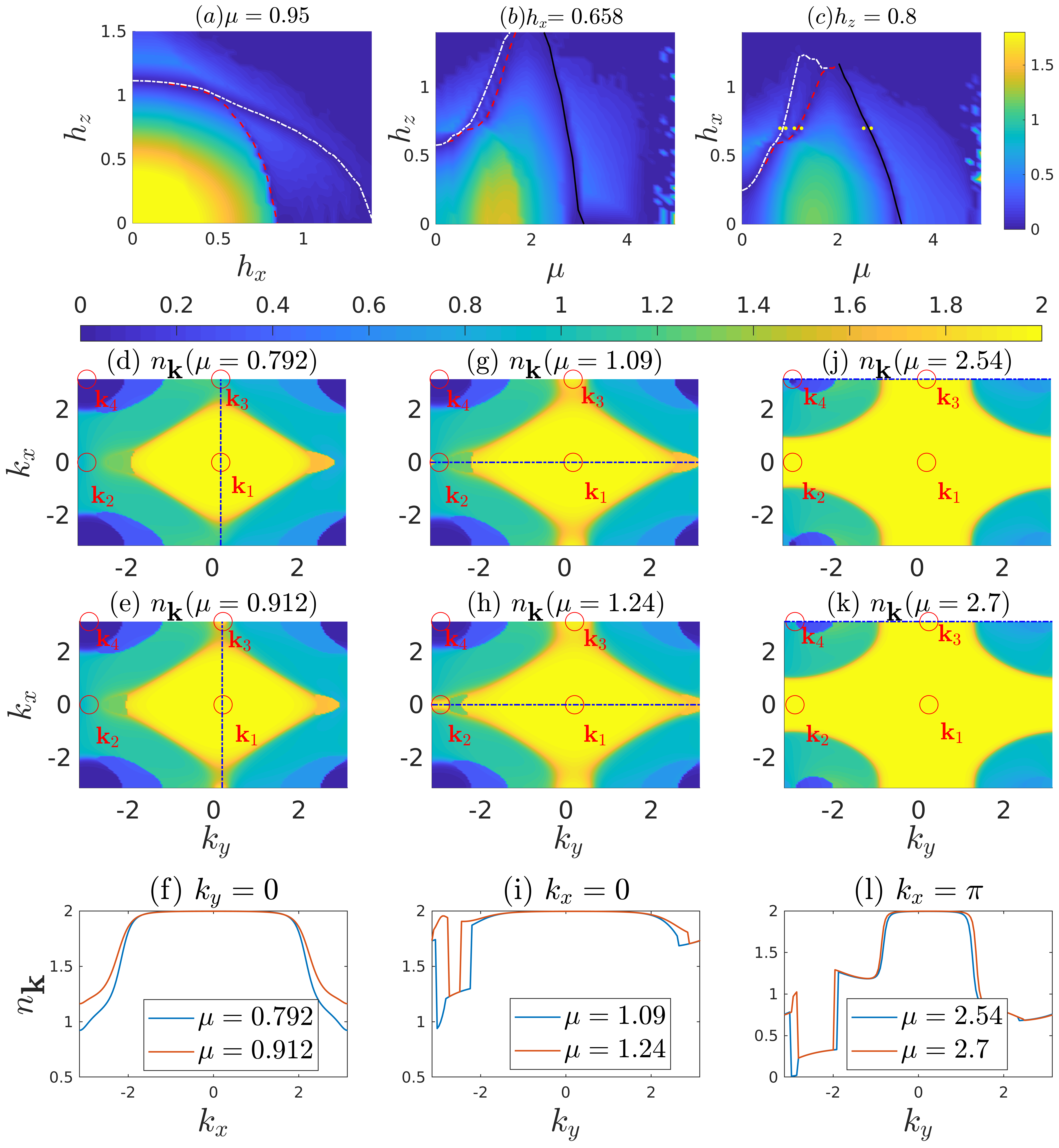}
\caption{\label{fig:4}(a)-(c) The minimum energy gap $E_g$  for $(h_x,h_z)$, $(\mu,h_z)$ and $(\mu,h_x)$-phase diagrams, respectively, shown above in figures~\ref{fig:1}(b), \ref{fig:3}(a) and \ref{fig:3}(c). Red, white and black lines correspond to analytical gap closing condition equations at $\bk_2 = (0,-\pi + \tilde{q}_y/2)$, $\bk_3 = (\pi,\tilde{q}_y/2)$ and $\bk_4 = (\pi,-\pi + \tilde{q}_y/2)$, respectively. Numerically and analytically computed gap closings are in a good agreement with the topological phase diagrams shown above. (d)-(l) Momentum density distributions $n_\bk$ for $\mu = 0.792$, $\mu = 0.912$, $\mu = 1.09$, $\mu = 1.24$, $\mu = 2.54$ and $\mu = 2.7$, corresponding to the six yellow dots shown in (c). Panels in two upper rows present $n_\bk$ in the first Brillouin zone and the lowest panels depict $n_\bk$ along the blue dash-dotted lines plotted in the upper panels. Furthermore, the red open circles in the upper panels indicate the locations of the possible gap closing momenta $\bk_1$, $\bk_2$, $\bk_3$ and $\bk_4$.}
\end{figure*}

\subsection{Topological phase transitions}

Topological phase diagrams presented here and in \cite{guo:2018} for a square lattice are relatively rich compared to the topological phase diagrams of Rashba-coupled 2D continuum where they are characterized by $C=1$ only. This can be explained by considering possible topological phase transitions which occur when the bulk energy gap $E_g$ between the quasi-particle eigenvalues $E^+_{\bk,\nu}$ and quasi-holes $E^-_{\bk,\nu}$ closes and reopens. Because of the intrinsic particle-hole symmetry present in our system, topological phase transitions can occur when the gap closes and reopens in particle-hole symmetric points \cite{ghosh:2010}. In continuum there exists only one particle-hole symmetric point, i.e. $\textbf{k} = (k_x,k_y) = (0,\tilde{q}_y/2)$. However, in a square lattice there are four different particle-hole symmetric points, namely $\bk_1 = (0,\tilde{q}_y/2)$, $\bk_2 = (0,-\pi + \tilde{q}_y/2)$, $\bk_3 = (\pi,\tilde{q}_y/2)$ and $\bk_4 = (\pi,-\pi + \tilde{q}_y/2)$ which yields four different gap closing equations instead of only one. Therefore, it is reasonable to find more distinct topological phases in a lattice system than in continuum. For similar reasons, topological phase diagrams studied in \cite{guo:2017} in case of triangular lattices possessed many distinct topological states characterized by different Chern numbers. Analytical gap-closing equations for the square lattice geometry are provided in appendix~\ref{app:gaps}.

In figures~\ref{fig:4}(a)-(c) we plot the minimum energy gap $E_g$ for $(h_x,h_z)$, $(\mu,h_z)$ and $(\mu,h_x)$-phase diagrams shown previously in figures~\ref{fig:1}(b), \ref{fig:3}(a) and (c). One can see that $E_g$ goes to zero at the topological phase boundaries as expected. In figures~\ref{fig:4} (a)-(c) we also depict the fulfilled analytical gap closing conditions which match with numerically computed topological boundaries. Analytical gap closing conditions can be thus used to identify distinct topological transitions in terms of the gap closing locations in the momentum space.

From figures~\ref{fig:4}(a)-(c) we see that the Chern invariant changes by one when the gap closes in one of the particle-hole symmetric momenta. However, when the system enters from the trivial $C=0$  phase to $C=-2$ phase, the gap closes simultaneously in two different momenta. This is consistent with the theory presented in \cite{ghosh:2010} considering the connection between the Chern number and gap closings at particle-hole symmetric points: if the Chern number changes by an even (odd) number at a topological phase transition, then the number of gap-closing particle-hole symmetric momenta is even (odd). %For more details, see \cite{ghosh:2010}.

We further investigate the topological phase transitions in figures~\ref{fig:4}(d)-(l), where we present the momentum density distributions $n_\textbf{k} = n_{\uparrow\textbf{k}} + n_{\downarrow\textbf{k}} = \langle c^\dag_{\uparrow \bk} c_{\uparrow \bk} \rangle + \langle c^\dag_{\downarrow \bk} c_{\downarrow \bk} \rangle$ for six different values of $\mu$, corresponding to six yellow dots depicted in figure~\ref{fig:4}(c). The topological transition corresponding to the gap closing at $\bk_3$ is studied in figures~\ref{fig:4}(d)-(e), and correspondingly closings at $\bk_2$ and $\bk_4$ are investigated in figures~\ref{fig:4}(g)-(i) and figures~\ref{fig:4}(j)-(l), respectively. 

By comparing the momentum distributions in figures~\ref{fig:4}(d)-(e) shown for $\mu = 0.792$ and $\mu = 0.912$, we observe that once the system goes through the topological transition identified by the gap closing and reopening at $\bk_3$ [white line in figure~\ref{fig:4}(c)], the momentum distribution changes qualitatively in the vicinity of $\bk_3$. This is further shown in figure~\ref{fig:4}(f) where $n_\bk$ for both cases is plotted at $k_y = 0$ along the blue dash-dotted line depicted in figures~\ref{fig:4}(d)-(e). In a similar fashion, one sees from figures~\ref{fig:4}(g)-(i) that the topological transition corresponding to the gap closing at $\bk_2$ [red line in figure~\ref{fig:4}(c)] is identified as an emergence of a prominent density peak around $\bk_2$ as clearly illustrated in figure~\ref{fig:4}(i). A similar peak can be also observed for the topological transition corresponding to $\bk_4$ though less pronounced as shown in figures~\ref{fig:4}(j)-(l). 

Drastic qualitative changes in the momentum distributions at the topological phase boundaries imply that one could experimentally measure and distinguish different topological phases and phase transitions in ultracold gas systems by investigating the total density distributions with the time-of-flight measurements. A similar idea to measure topological phase transitions were proposed in \cite{zhang:2013} in case of a simpler continuum system. Our findings show that density measurements could be applied also in lattice systems to resolve different topological phases.

\begin{figure}
\includegraphics[width=0.9\columnwidth]{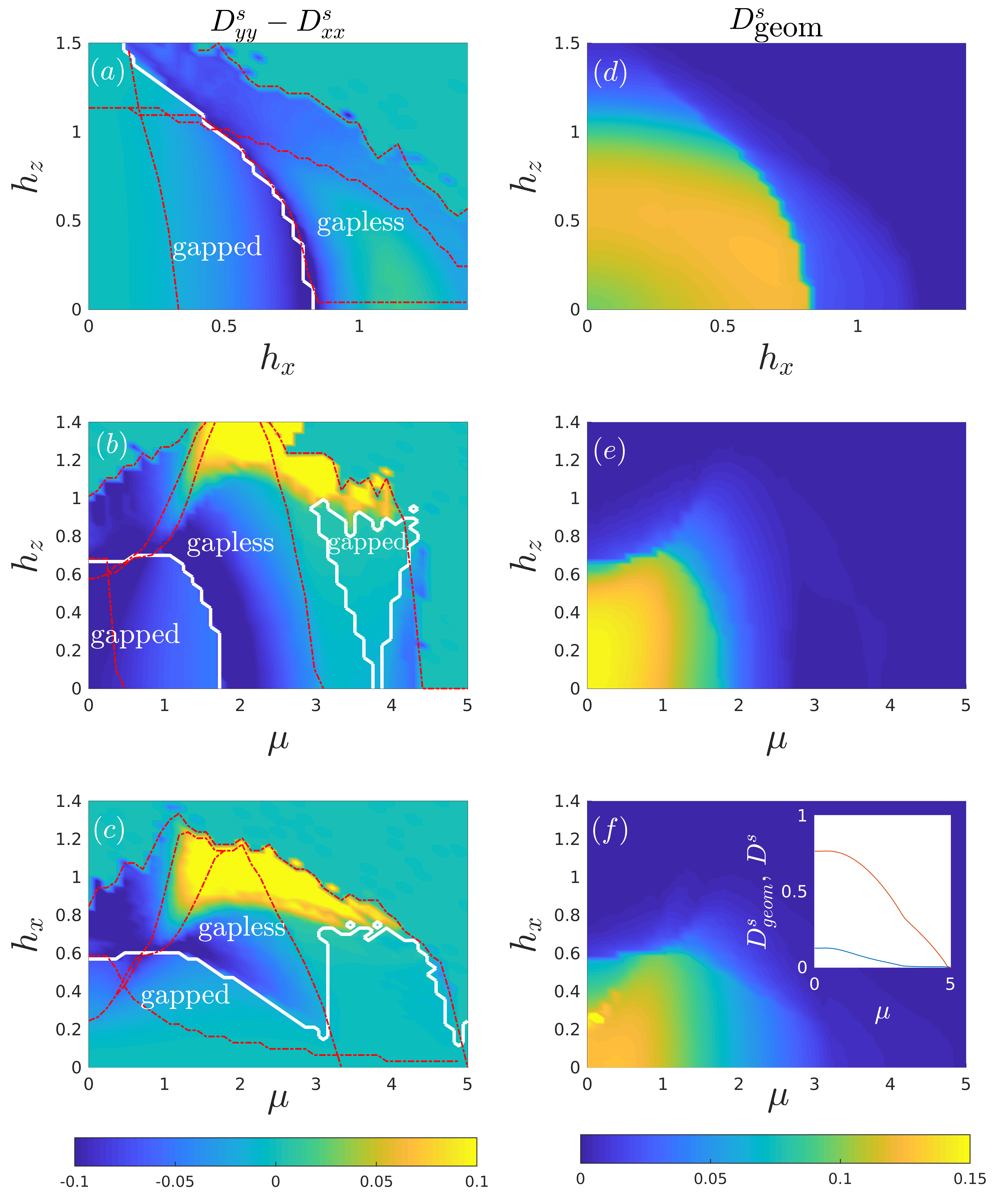}
\caption{\label{fig:5}(a)-(c) The difference of perpendicular superfluid weight components $D^s_{\textrm{diff}} = D^s_{yy} - D^s_{xx}$ for $(h_x,h_z)$, $(\mu,h_z)$ and $(\mu,h_x)$-phase diagrams, respectively. The white solid lines depict the boundaries between the gapped and gapless superfluid states. The red dash-dotted lines correspond to phase boundaries shown in figures~\ref{fig:1} and \ref{fig:3}. (d)-(f) The geometric contribution $D^s_{\textrm{geom}}$ for $(h_x,h_z)$, $(\mu,h_z)$ and $(\mu,h_x)$-phase diagrams. The inset in (f) shows the total superfluid weight $D^s$ (red line) and $D^s_{\textrm{geom}}$ (blue line) for $h_x =0$. In all three cases the geometric contribution is smaller than the total superfluid weight and more or less vanishes when the system enters the large-$\tilde{q}_y$ FF regime.}
\end{figure}

\subsection{Components of the superfluid weight}

As the single particle energy dispersions are deformed in the $y$-direction but not in the $x$-direction, the rotational symmetry of the lattice is broken. This manifests itself as different superfluid weight components in the $x$- and $y$-directions, i.e. $D^s_{xx} \neq D^s_{yy}$. As the Cooper pair momentum is in the $y$-direction, we call $D^s_{yy}$ as the longitudinal and $D^s_{xx}$ as the transverse component. Because $D^s_{xx} \neq D^s_{yy}$, the system has different current response in these directions when exposed to an external magnetic field. Therefore, it is meaningful to investigate the difference of the longitudinal and transverse components, $D^s_{\textrm{diff}} \equiv D^s_{yy} - D^s_{xx}$, to see how it behaves as a function of our system parameters. We focus only on the diagonal elements of $D^s$ as the off-diagonal elements in our case are always zero, i.e. $D^s_{xy} = D^s_{yx} = 0$.

In figures~\ref{fig:5}(a)-(c) we present $D^s_{\textrm{diff}}$ for $(h_x,h_z)$, $(\mu,h_z)$ and $(\mu,h_x)$-phase diagrams, respectively, shown above in figures~\ref{fig:1}(b), \ref{fig:3}(a) and \ref{fig:3}(c). In  all three cases, $D^s_{\textrm{diff}}$ more or less vanishes in large parts of the phase diagrams. However, especially when entering the large-$\tilde{q}_y$ FF region from the small-$\tilde{q}_y$ region, $D^s_{\textrm{diff}}$ reaches local minima and becomes negative. On the other hand, from figures~\ref{fig:5}(b)-(c) we see that there also exists a parameter region where $D^s_{\textrm{diff}}$ is positive and that the tFF$_{-2}$-phase in figure~\ref{fig:5}(c) near half-filling is clearly distinguishable from the neighboring phases. Therefore, by measuring $D^s_{\textrm{diff}}$ one could in principle distinguish some of the phase transitions existing in the system. It is interesting to note that,  in the presence of SOC, the transverse component can be larger than the longitudinal component, in contrast to 2D continuum where the absence of SOC results in the vanishing transverse component and thus the vanishing BKT temperature $T_{BKT} =0$ \cite{yin:2014}.

In addition to $D^s_{\textrm{diff}}$, in figures~\ref{fig:5}(a)-(c) we also plot with solid white lines the boundaries of gapped and gapless superfluid phases. Consistent with previous literature \cite{zhang:2013,qu:2013,cao:2014,cao:2015}, we call the system gapless (or nodal) if one or more of the Bogoliubov quasi-hole branches reach the zero-energy in some part of the momentum space, i.e. the quasi-particle excitation energy vanishes for some momenta. Note that this does not (necessarily) mean that the topological energy gap $E_g$ closes as $E_g$ is the difference of the highest quasi-hole and the lowest quasi-particle energy at the same momentum $\bk$ such that both are also the eigenvalues of $\mathcal{H}_\bk$, whereas the highest quasi-hole and the lowest quasi-particle energy are not necessarily at the same momentum.

From figures~\ref{fig:5}(a) and (c) we see that the system stays gapped at low in-plane Zeeman field strengths which is consistent with continuum results \cite{cao:2014,cao:2015}. For larger $h_x$ the system becomes eventually gapless and one can observe topologically trivial and non-trivial nodal FF phases. By comparing figures~\ref{fig:1}(b), \ref{fig:3}(a) and \ref{fig:3}(c) to figures~\ref{fig:5}(a)-(c) we can make a remark that FF states with small momenta $\tilde{q}_y$ are gapped. Furthermore, we observe from figures~\ref{fig:5}(a)-(c) that the transitions between the gapped and gapless states at moderate Zeeman fields and chemical potentials coincide with the prominent minima of  $D^s_{\textrm{diff}}$. This is consistent with the findings of \cite{cao:2015} where it was shown that the longitudinal component exhibits a clear minimum when the system becomes gapless. However, in figures~\ref{fig:5}(b)-(c) we see the system reaching a gapped region again at large enough $\mu$ without such a drastic change of $D^s_{\textrm{diff}}$ than at smaller values of $\mu$.

In addition to different spatial components, one can also investigate the role of the geometric superfluid weight contribution $D^s_{\mathrm{geom}}$ which is presented for $(h_x,h_z)$, $(\mu,h_z)$ and $(\mu,h_x)$-phase diagrams in figures~\ref{fig:5}(d)-(f). We see that for BCS states and gapped FF states of small Cooper pair momenta,  the geometric contribution is notable but is otherwise vanishingly small. In all the cases the geometric contribution is relatively small compared to the total superfluid weight $D^s$ which is, as an example, illustrated in the inset of figure~\ref{fig:5}(f) where $D^s_{\mathrm{geom}}$ and $D^s$ are both plotted for $h_x = 0$. At largest, the geometric contribution is responsible up to $18$ percent of the total superfluid weight which is fairly similar to what was reported in \cite{iskin:2017}, where the geometric part was found to contribute up to a quarter of the total superfluid weight in case of a spin-orbit-coupled 2D BCS continuum model. In more complicated multiband lattices, such as honeycomb lattice or Lieb lattice (which also possesses a flat band), the geometric contribution in the presence of SOC might be more important than in our simple square lattice example as the geometric contribution is intrinsically a multiband effect \cite{peotta:2015}. 

\section{Conclusions and outlook}\label{section_four}
In this work we have investigated the stability of exotic FF superfluid states in a lattice system by computing the superfluid weight and BKT transition temperatures systematically for various system parameters. The derivation of the superfluid weight is based on the linear response theory and is an extension of the previous studies of \cite{liang:2017,peotta:2015} where only BCS ansatzes without spin-flipping terms were considered. Our method applies to BCS and FF states in the presence of arbitrary spin-flipping processes and lattice geometries. We find that, as previously in case of conventional BCS theory without the spin-flipping contribution, also in case of FF phases and with spin-flipping terms one can divide the total superfluid weight to conventional and geometric superfluid contributions. 

We have focused on a square lattice geometry in the presence of the Rashba-coupling. One of the main findings of this article is that conventional spin-imbalance-induced FF states, in the absence of SOC, indeed have finite BKT transition temperatures in a lattice geometry. For our parameters they could be observed at $T\sim 0.1t$. In earlier theoretical studies it has been predicted that FF states could exist in two-dimensional lattice systems \cite{kinnunen:2018,baarsma:2016,koponen:2007,gukelberger:2016} but the stability in terms of the BKT transition has never been investigated in lattice systems. By computing $T_{BKT}$ we show that two-dimensional FFLO superfluids should be realizable in finite temperatures. By applying SOC, we show that FF states in a lattice can be further stabilized and for our parameter regime BKT temperatures as high as $T \sim 0.17-0.3t$ can be reached. Spin-orbit coupling also enables the existence of topological nodal and gapped FF states, for which we show the BKT transitions to occur at highest around $T_{BKT} \sim 0.15t$. 

For literature comparison, we estimated that $T_{BKT} \approx 0.25t$ at $U=-4t$  for usual spin-balanced BCS state at half-filling without SOC, see figure~\ref{fig:1}(c), whereas in \cite{paiva:2004} the corresponding estimate obtained by Monte Carlo simulations was $T_{BKT} \sim 0.10-0.13t$.  Thus, our mean-field approach probably overestimates $T_{BKT}$ in case of a simple square lattice. However, in \cite{julku:2016,liang:2017} the superfluid weights of BCS states, derived in the framework of mean-field theory, were shown to agree reasonably well with more sophisticated theoretical methods in case of multiband systems. Thus, it is expected that our mean-field superfluid equations are in better agreement with beyond-mean-field methods when considering multiband lattice models.

We have also shown that different topological FF phases and phase transitions could be observed by investigating the total momentum density profiles. When the system goes through a topological phase transition, the momentum distribution develops peaks or dips in the vicinity of momenta in which the energy gap closes and re-opens. In addition to density distributions, also the relative behavior of the longitudinal and transverse superfluid weight components yields implications about the phase transitions, especially near the boundaries of gapless and gapped superfluid phases. Therefore, our work paves the way for stabilizing and identifying exotic topological FF phases in lattice systems.

In future studies it would be interesting to see how stable FF states are in multiband models. This could be investigated straightforwardly with our superfluid weight equations as they hold for an arbitrary multiband system. Especially intriguing could be systems which possess both dispersive and flat bands such as kagome or Lieb lattices. In these systems the conventional spin-imbalanced FF states were recently shown to exhibit exotic deformation of Fermi surfaces due to the presence of a flat band \cite{huhtinen:2018}. In multiband systems one could also expect the geometric superfluid contribution to play a role, in contrast to our square lattice system where the geometric contribution was only non-zero for BCS and gapped FF phases. Furthermore, in flat band systems mean-field theory is shown to be in good agreement with more advanced beyond mean-field approaches \cite{liang:2017,julku:2016,tovmasyan:2016}. Flat band systems are tempting also because it is expected that their superfluid transition temperatures in the weak-coupling region are higher than in dispersive systems \cite{kopnin:2011,heikkila:2011,peotta:2015,liang:2017,julku:2016} and thus they could provide a way to realize exotic FFLO phases at high temperatures.

\appendix
\section{Details on deriving the superfluid weight}
\label{app:derivation}
Here we briefly go through how one obtains the final form for the superfluid weight $D^s$ shown in \eqref{sfw} from the intermediate result \eqref{ds_int}. As one can see from \eqref{ds_int}, there exists two terms in $K_{\mu\nu}$, the first being the diamagnetic and the second one the paramagnetic contribution, $K_{\mu\nu,\textrm{dia}}$, $K_{\mu\nu,\textrm{para}}$, respectively. We focus on the diamagnetic term and after that just give the result for the paramagnetic term as the derivation for both terms is essentially the same. 

In the diamagnetic term there exists a double derivative $\partial_\mu \partial_\nu \mathcal{H}(\bk)$ which can be transformed to a single derivative via integrating by parts:
\begin{align}
\label{app1}
K_{\mu\nu,\textrm{dia}} &=\frac{1}{\beta} \sum_{\bk,\Omega_m} \textrm{Tr} \Big[\partial_\mu \partial_\nu \mathcal{H}(\bk) P_+ G(i\Omega_m,\bk)\Big] \nonumber \\ &= - \frac{1}{\beta} \sum_{\bk,\Omega_m}  \textrm{Tr} \Big[ \partial_\mu \mathcal{H}(\bk) P_+ \partial_\nu G(i\Omega_m,\bk) \Big].
\end{align}
Because $G(i\Omega_m,\bk) = 1/(i\Omega_m - \mathcal{H}(\bk))$, we have  $\partial_\nu G^{-1} = -\partial_\nu \mathcal{H}$ and because $\partial_\nu(GG^{-1}) = 0$ we also have $\partial_\nu G = -G \partial_\nu G^{-1} G$ so that \eqref{app1} can be written as
\begin{align}
K_{\mu\nu,\textrm{dia}} &= - \frac{1}{\beta} \sum_{\bk,\Omega_m} \textrm{Tr} \Big[ \partial_\mu \mathcal{H}(\bk) P_+ G(i\Omega_m,\bk) \partial_\nu \mathcal{H}(\bk) G(i\Omega_m,\bk) \Big] \nonumber \\
&= - \frac{1}{\beta} \sum_{\bk,\Omega_m}\sum_i^{4M}\langle \phi_i(\bk) | \partial_\mu \mathcal{H}(\bk) P_+ G(i\Omega_m,\bk) \partial_\nu \mathcal{H}(\bk) G(i\Omega_m,\bk) |\phi_i(\bk) \rangle,
\end{align}
where $|\phi_i(\bk) \rangle$ are the eigenvectors of $\mathcal{H}_\bk$. By using the completeness relation $\sum_j |\phi_j(\bk) \rangle \langle \phi_j(\bk) | = 1$ and the alternative form for $G(i\Omega_m,\bk)$
\begin{align}
G(i\Omega_m,\bk) = \sum_{l=1}^{4M}\frac{|\phi_l(\bk) \rangle \langle \phi_l(\bk) |}{i\Omega_m - E_{l,\bk}}
\end{align}
we obtain 
\begin{align}
K_{\mu\nu,\textrm{dia}} =& - \frac{1}{\beta} \sum_{\bk,\Omega_m}\sum_{i,j}^{4M}\langle \phi_i(\bk) | \partial_\mu \mathcal{H}(\bk) P_+ | \phi_j(\bk) \rangle \nonumber \\ &\times \langle \phi_j(\bk) | \partial_\nu \mathcal{H}(\bk)| \phi_i(\bk) \rangle \frac{1}{(i\Omega_m - E_{j,\bk})(i\Omega_m - E_{i,\bk})}.
\end{align}
The summation over the Matsubara frequencies $\Omega_m$ can be carried out analytically yielding
\begin{align}
K_{\mu\nu,\textrm{dia}} = &\sum_{\bk,ij} \langle \phi_i(\bk) | \partial_\mu \mathcal{H}(\bk) P_+ | \phi_j(\bk) \rangle \nonumber \\ &\times \langle \phi_j(\bk) | \partial_\nu \mathcal{H}(\bk)| \phi_i(\bk) \rangle \frac{n(E_{j,\bk})-n(E_{i,\bk})}{E_{i,\bk} - E_{j,\bk}}.
\end{align}
In a similar fashion one derives the following result for the paramagnetic term:
\begin{align}
K_{\mu\nu,\textrm{para}}(\bq \rightarrow 0, 0) = &-\sum_{\bk,ij} \langle \phi_i(\bk) | \partial_\mu \mathcal{H}(\bk)P_+ | \phi_j(\bk) \rangle \nonumber \\ &\times \langle \phi_j(\bk) | \partial_\nu \mathcal{H}(\bk) \hat{\gamma}_z| \phi_i(\bk) \rangle \frac{n(E_{j,\bk})-n(E_{i,\bk})}{E_{i,\bk} - E_{j,\bk}}.
\end{align}
As $D^s_{\mu\nu} = K_{\mu\nu}(\bq \rightarrow 0,0) = K_{\mu\nu,\textrm{dia}} + K_{\mu\nu,\textrm{para}}(\bq \rightarrow 0, 0)$ and $P_- =  (I_{4M} - \hat{\gamma}_z)/2$, one readily obtains the final result presented in \eqref{sfw}.

\section{Geometric contribution of the superfluid weight}
\label{app:a}
In this appendix we show how the total superfluid weight $D^s$  presented in \eqref{sfw} can be split to the so-called conventional and geometric contributions, $D^s_{\textrm{conv}}$ and $D^s_{\textrm{geom}}$. We start by expressing the eigenvectors $| \phi_i(\bk) \rangle$ of $\mathcal{H}(\bk)$ in terms of the eigenvectors of $\mathcal{H}_p(\bk)$ and $\mathcal{H}_h(\bk)$ as follows
\begin{align}
&|\phi_i(\bk) \rangle = \sum_{m=1}^{2M}\Big( w_{p,im}   |+\rangle \otimes |m\rangle^p  +  w_{h,im}  |-\rangle \otimes |m \rangle^h \Big),
\end{align}
where $|m\rangle^p$ ( $|m\rangle^h$) are the eigenvectors of $\mathcal{H}_p$ ($\mathcal{H}_h$) and $|\pm \rangle$ are the eigenvectors of $\hat{\sigma_z} \otimes I_{2M}$ with the eigenvalues $\pm 1$. By noting that 
\begin{align}
\partial_\mu \mathcal{H}(\bk) = \begin{bmatrix}
\partial_\mu \mathcal{H}_p(\bk) & 0 \\
0 & - \partial_\mu \mathcal{H}_h (\bk - \bqt)
\end{bmatrix}
\end{align}
we can rewrite \eqref{sfw} as
\begin{align}
\label{sfw2}
D^s_{\mu\nu} = &\sum_{\bk,ij} \frac{n(E_j) - n(E_i)}{E_i - E_j} \nonumber \\ & \times\sum_{m_1,m_2}^{2M}\Big[ w^*_{p,im_1} w_{p,jm_2} {}^p \langle m_1 | \partial_\mu \mathcal{H}_p(\bk) | m_2 \rangle^p  \Big] \nonumber \\
& \times \sum_{m_3,m_4}^{2M} \Big[ w^*_{h,jm_3} w_{h,im_4}  {}^h \langle m_3 | -\partial_\nu \mathcal{H}_h(\bk - \tilde{\bq}) | m_4 \rangle^h \Big] \nonumber \\
=& \sum_{\substack{\bk \\m_1,m_2,\\m_3,m_4}} W_{m_1m_2}^{m_3m_4} \big({}^p \langle m_1 | \partial_\mu \mathcal{H}_p | m_2 \rangle^p {}^h \langle m_3 | -\partial_\nu\mathcal{H}_h | m_4 \rangle^h \big),
\end{align}
where 
\begin{align}
W_{m_1m_2}^{m_3m_4} = \sum_{ij} \frac{n(E_j) - n(E_i)}{E_i - E_j} w^*_{p,im_1}w_{p,jm_2}w^*_{h,jm_3}w_{h,im_4}.
\end{align}
and
\begin{align}
\label{sfw3}
{}^p \langle m_1 | \partial_\mu \mathcal{H}_p | m_2 \rangle^p =& \delta_{m_1,m_2} \epsilon_{m_1} + (\epsilon_{m_1} - \epsilon_{m_2}) ~ {}^p \langle \partial_\mu m_1 | m_2 \rangle^p .
\end{align}
Here $\epsilon_{m_i}$ are the eigenvalues for $\mathcal{H}_p$. Similar expression holds also for the ${}^h \langle m_3 | -\partial_\nu\mathcal{H}_h | m_4 \rangle^h$ elements. From \eqref{sfw2}-\eqref{sfw3} we note that there exists two superfluid weight components. The component which is called the conventional contribution $D^s_{\textrm{conv}}$ consists of matrix elements with $m_1 = m_2$  and $m_3 = m_4$. As can be seen from \eqref{sfw3}, the conventional contribution depends only on the single-particle dispersions $\epsilon_{m_i}$. The remaining part is the geometric contribution $D^s_{\textrm{geom}}$ and it depends on the geometric properties of the Bloch functions, $| m_i \rangle^p$ and $| m_i \rangle^h$.

\section{Comparison of the superfluid weight and the BKT temperature to previous literature}
\label{app:benchmark}
As our equations for the superfluid weight hold for arbitrary geometries in the presence and absence of SOC, we can make direct comparisons to previous studies. As the first benchmark, we reproduced the superfluid weight results of \cite{julku:2016} where BCS states in the Lieb lattice geometry without the SOC are studied by applying mean-field theory and exact diagonalization (ED) methods. One should emphasize that mean-field equations used in \cite{julku:2016} to compute the superfluid weight are derived by not using the linear response theory as in our study but by using an alternative approach based on the definition given in \cite{peotta:2015}. Our method yields exactly the same results as the alternative mean-field and ED approaches of \cite{julku:2016}. Furthermore, we have checked that in the continuum limit our expression for the superfluid weight reduces to the expressions presented in \cite{iskin:2017} where BCS states in spin-orbit-coupled 2D continuum were considered.

We also benchmarked our equations by computing $T_{BKT}$ in case of BCS phases for a 2D square lattice geometry with the same parameters that were used in \cite{yajie:2016} where topological BCS states in the presence of the SOC were studied. With our equations we find the same functional behavior for $T_{BKT}$  as a function of $U$  but our results are exactly a factor of two larger than those presented in \cite{yajie:2016}. The reason for this difference is because in \cite{yajie:2016}, the phase fluctuations of the order parameter are rescaled by a factor of $1/\sqrt{2}$ [see equation~(33) in \cite{yajie:2016}]. With this rescaling, the periodicity of the $\phi$ field in (38) becomes $2\sqrt{2}\pi$ and therefore the expression for the BKT transition temperature [equation~(39)] should be multiplied by a factor of 2.

\section{Analytic equations for the gap closing and reopening conditions}
\label{app:gaps}
In this appendix we show the analytical equations that were used to depict the topological phase transitions in figure~\ref{fig:4}. The energy gap $E_g$ between the quasi-particle eigenvalues $E^+_{\bk,\nu}$ and quasi-holes $E^-_{\bk,\nu}$ can only close and reopen at particle-hole-symmetric points which in our case are $\bk_1 = (0,\tilde{q}_y/2)$, $\bk_2 = (0,-\pi + \tilde{q}_y/2)$, $\bk_3 = (\pi,\tilde{q}_y/2)$ and $\bk_4 = (\pi,-\pi + \tilde{q}_y/2)$. The single-particle Hamiltonian $\mathcal{H}_p$ in these four points can be diagonalized analytically which yields four eigenvalues, namely $E^-_{\bk,1} \leq E^-_{\bk,2} \leq E^+_{\bk,2} \leq E^+_{\bk,1}$. By demanding $E^-_{\bk,2} = E^+_{\bk,2}$ at each of the four particle-hole symmetric momenta, one obtains the four gap closing equations which read
\begin{align}
h_z^2 =& 6 + \Delta^2 + 4\mu + \mu^2 + 4(2+\mu)\cos(\tilde{q}_y/2) + 2\cos(\tilde{q}_y) - h_x^2 +2\lambda^2[\cos(\tilde{q}_y) -1] \nonumber \\
& +4h_x\lambda\sin(\tilde{q}_y/2) \\
h_z^2 =& 6 + \Delta^2 + 4\mu + \mu^2 - 4(2+\mu)\cos(\tilde{q}_y/2) + 2\cos(\tilde{q}_y) - h_x^2 +2\lambda^2[\cos(\tilde{q}_y) -1] \nonumber \\
& -4h_x\lambda\sin(\tilde{q}_y/2) \\
h_z^2 =& 6 + \Delta^2 - 4\mu + \mu^2 + 4(2+\mu)\cos(\tilde{q}_y/2) + 2\cos(\tilde{q}_y) - h_x^2 +2\lambda^2[\cos(\tilde{q}_y) -1] \nonumber \\
& +4h_x\lambda\sin(\tilde{q}_y/2) \\
h_z^2 =& 6 + \Delta^2 - 4\mu + \mu^2 - 4(2+\mu)\cos(\tilde{q}_y/2) + 2\cos(\tilde{q}_y) - h_x^2 +2\lambda^2[\cos(\tilde{q}_y) -1] \nonumber \\
& -4h_x\lambda\sin(\tilde{q}_y/2).
\end{align}
By solving these equations for different values of $h_x$, $h_z$ and $\mu$, one obtains the topological boundaries shown in figures~\ref{fig:4}(a)-(c).

\section{Direction of the Cooper pair momentum}
\label{app:qbm}
\begin{figure}
\includegraphics[width=1.0\columnwidth]{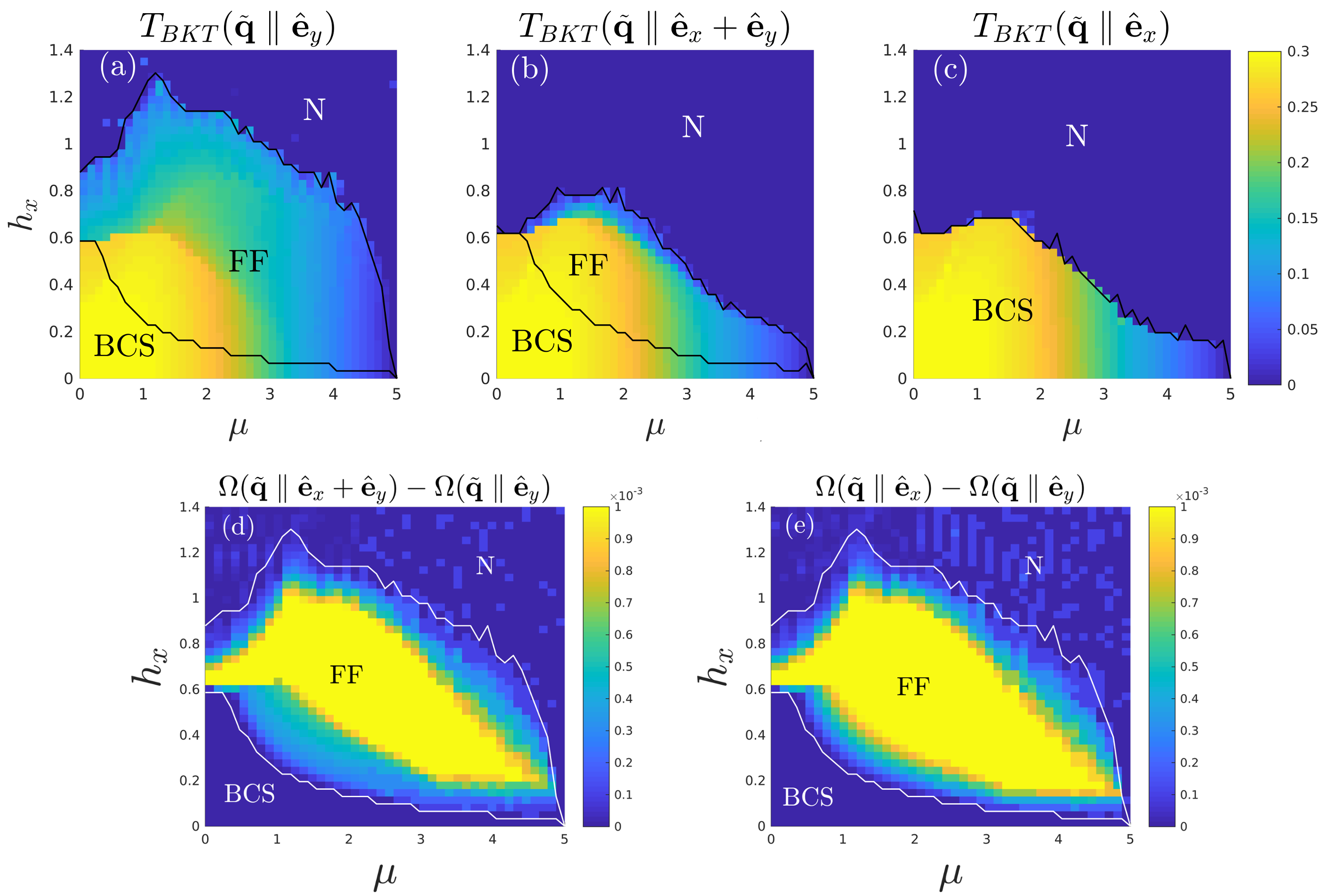}
\caption{\label{fig:appendix}(a)-(c) Computed phase diagrams as functions of $\mu$ and $h_x$ by assuming $\tilde{\textbf{q}}\parallel \hat{\textbf{e}}_y$ (a), $\tilde{\textbf{q}}\parallel \hat{\textbf{e}}_x+\hat{\textbf{e}}_y$ (b) and $\tilde{\textbf{q}}\parallel \hat{\textbf{e}}_x$ (c). Black solid lines depict the phase boundaries between BCS, FF and normal states. (d)-(e) Grand canonical thermodynamic potential differences between the cases $\tilde{\textbf{q}}\parallel \hat{\textbf{e}}_x+\hat{\textbf{e}}_y$ and $\tilde{\textbf{q}}\parallel \hat{\textbf{e}}_y$ (d), and between $\tilde{\textbf{q}}\parallel \hat{\textbf{e}}_x$ and $\tilde{\textbf{q}}\parallel \hat{\textbf{e}}_y$ (e). White lines are the phase boundaries in case of $\tilde{\textbf{q}}\parallel \hat{\textbf{e}}_y$.}
\end{figure}

In our computations the Cooper pair momentum $\tilde{\bq}$ is in the $y$-direction, i.e. $\tilde{\bq}\parallel \hat{\textbf{e}}_y$, consistent with earlier studies concerning lattice systems \cite{xu:2014,guo:2018,guo:2017}. We have extensively tested numerically that indeed the wavevector in the $y$-direction minimizes the thermodynamic potential with and without SOC for all the used input parameters. As an example, we have demonstrated this in figure \ref{fig:appendix}. In figures \ref{fig:appendix}(a)-(c) we plot the $(\mu,h_x)$-phase diagram for three different cases: in (a) the thermodynamic potential $\Omega$ is minimized so that $\tilde{\textbf{q}}$ is taken to be in the $y$-direction, in (b) $\tilde{\textbf{q}}$ is along the diagonal direction ($\tilde{q}_x = \tilde{q}_y$) and in (c) $\tilde{\textbf{q}}$ is in the $x$-direction. The out-of-plane Zeeman field is chosen to be $h_z = 0.8$, the spin-orbit-coupling is $\lambda = 0.75$ and the interaction strength is $U=-4$ so the phase diagram in figure \ref{fig:appendix}(a) is the same as in figure \ref{fig:3}(c) in the main text. We see how gradually the FF region becomes smaller when the wavevector is forced to deviate from the $y$-direction. In figures \ref{fig:appendix}(d)-(e) we compare the thermodynamic potentials $\Omega$ of these three different cases. In figure \ref{fig:appendix}(d) the thermodynamic potential difference of cases $\tilde{\textbf{q}}\parallel\hat{\textbf{e}}_x + \hat{\textbf{e}}_y$ and $\tilde{\textbf{q}}\parallel\hat{\textbf{e}}_y$  is plotted and correspondingly in figure \ref{fig:appendix}(e) the thermodynamic potential difference of cases $\tilde{\textbf{q}}\parallel\hat{\textbf{e}}_x$ and $\tilde{\textbf{q}}\parallel\hat{\textbf{e}}_y$ is depicted. White lines show the phase boundaries between the BCS, FF and normal phases in case of $\tilde{\textbf{q}}\parallel\hat{\textbf{e}}_y$. We see that within the BCS phase the thermodynamic potential is the same regardless of the direction of the wavevector as in the BCS phase the Cooper pair momentum is zero. When entering the FF phase, it is clear that phase diagrams shown in figures \ref{fig:appendix}(b)-(c) do not depict the true ground states as their thermodynamic potentials are higher than in case of $\tilde{\textbf{q}}\parallel\hat{\textbf{e}}_y$. Thus the states shown in figure \ref{fig:appendix}(a) with $\tilde{\textbf{q}}\parallel\hat{\textbf{e}}_y$ are energetically more stable than the states with the Cooper pair momentum in the diagonal or $x$-direction.

In figure \ref{fig:appendix} we have only presented three different options for the direction of $\tilde{\textbf{q}}$ and only $(\mu,h_x)$-phase diagram. However, they represent the general trend of all the computations of our work: the thermodynamic potential reaches its minimum when $\tilde{\textbf{q}}$ is in the $y$-direction. We have confirmed this by choosing $20$ other directions between the $x$ and $y$-axes. Alternatively, we also  minimized the thermodynamic potential by letting $q_x$ and $q_y$ be independent parameters. As the thermodynamic potential can have many local minima as a function of $q_x$ and $q_y$, this procedure is not the most trustworthy for finding the global minimum. However, we did not find a single local minimum lying outside the $y$-axis that would have lower energy than the solutions we find by assuming $\tilde{\textbf{q}}$ $||$ $\hat{\textbf{e}}_y$. Therefore we are confident that our statements and results are correct within the mean-field theory framework.

\bibliography{bib_soc}

\end{document}